\documentclass[12pt]{iopart}
\pdfoutput=1
\usepackage{iopams}
\usepackage{ulem}
 
\expandafter\let\csname equation*\endcsname\relax
\expandafter\let\csname endequation*\endcsname\relax
\usepackage{amsmath,amssymb,mathtools}

\allowdisplaybreaks[1]

\usepackage[mathscr]{euscript}
\usepackage{url}
\usepackage{xcolor,hyperref,bm,graphicx}

\def\e{\operatorname{e}}
\renewcommand{\dh}{\partial}
\renewcommand{\d}{\mathrm{d}}

\renewcommand{\l}{\left}
\renewcommand{\r}{\right}
\newcommand{\defn}{\equiv}
\newcommand{\op}{\operatorname}
\newcommand{\bra}[1]{\l<#1\r|}
\newcommand{\ket}[1]{\l|#1\r>}
\newcommand{\braket}[2]{\l<#1|#2\r>}
\newcommand{\abs}[1]{\l|#1\r|}

\newcommand{\mean}[1]{\l<#1\r>}

\newcommand{\custombracket}[3]{\left#1\rule{0em}{#2}\right#3}
\newcommand{\inSet}[1]{\l\{#1\r\}}
\newcommand{\evalAt}[1]{\l.#1\r|}
\newcommand{\given}{\,|\,}
\newcommand{\LaplaceTransform}{\widetilde}
\newcommand{\la}{\langle}
\newcommand{\ra}{\rangle}

\newcommand{\lam}{\lambda}

\newcommand{\customHead}[1]{\bigskip\noindent {\boldmath \textbf{#1}}} 

\usepackage{cancel}

\begin{document}

\title{Local time of an Ornstein-Uhlenbeck particle}
\author{G. Kishore \footnote{kishoreg@iucaa.in}}
\address{Inter University Centre for Astronomy and Astrophysics, Post Bag 4, Ganeshkhind, Pune - 411007, India}
\author{Anupam Kundu \footnote{anupam.kundu@icts.res.in}}
\address{International Centre for Theoretical Sciences, Tata Institute of Fundamental Research, Bengaluru - 560089, India}

\begin{abstract}
In this paper, we study the local time spent by an Ornstein-Uhlenbeck particle at some location till time $t$. Using the Feynman-Kac formalism, the computation of the moment generating function of the local time can be mapped to the 
problem of finding the eigenvalues and eigenfunctions of a quantum particle. 

We employ quantum perturbation theory to compute the eigenvalues and eigenfunctions in powers of the argument of the moment generating function which particularly help to directly compute the cumulants and correlations among local times spent at different locations. In particular, we obtain explicit expressions of the mean, variance, and covariance of the local times  in the presence and in the absence of an absorbing boundary, conditioned on survival. In the absence of absorbing boundaries, we also study large deviations of the local time and compute exact asymptotic forms of the associated large deviation functions explicitly. 

In the second part of the paper, we extend our study of the statistics of local time of the Ornstein-Uhlenbeck particle to the case not conditioned on survival. In this case, one expects the distribution of the local time to reach a stationary distribution in the large time limit. Computations of such stationary distributions are known in the literature as the problem of first passage functionals. In this paper, we study the approach to this stationary state with time by providing a general formulation for evaluating the moment generating function. From this moment generating function, we compute the cumulants of the local time exhibiting the approach to the stationary values explicitly for a free particle and a Ornstein-Uhlenbeck particle. Our analytical results are verified and supported by numerical simulations.
\end{abstract}

\section{Introduction}
The local time  of a particle at a specified point in space is the total time spent by the particle at that location over a fixed duration of time 
\cite{Knight1969,Feller2008,Majumdar2005}. Local times measured at different locations together can describe the spatiotemporal properties of the motion of the particle. In the context of reaction-diffusion processes, this quantity can be a key controlling factor of the reaction \cite{Feller2008,Elgeti2015,Agmon1984,Berezhkovskii1998,Nguyen2010,Pal2019a,Greb1,Greb2,Greb3}. Usually, the reaction centres are heterogeneously distributed over space \cite{Grebenkov2006} and quite often their distribution fluctuates over time. In such cases, the rate or the yield of a particular reaction would highly depend on the amount of time a reaction agent spends at these reaction centres \cite{Wilemski1973,Benichou2005,Doi1975,Temkin1984}. These times are called residence times, and in the limit of small size of the reaction centres, the density of residence time spent is called the local time (density). In a slightly more complicated situation, one can think of a series of reactions which depend on each other. In such cases, the yield of these reactions may depend on the correlation between the residence times or local times spent at different  locations. 
For example, the time spent by a diffusing agent near a boundary or a catalytic surface plays a crucial role in describing various surface-mediated phenomena \cite{Greb1, Greb2, Greb3}.

In the physics and mathematics literature, local and residence times have been studied in detail in the context of Brownian functionals, both in presence and absence of external potentials, and with different types of boundary conditions \cite{Darling1957,Lamperti1958,Ito1974,Watanabe1995,Pitman3003,Borodin2002,Majumdar2005,Grebenkov2007}. The local time spent at position $y$ by a Brownian particle moving in one dimension over a time duration $t$, denoted by $\ell_t(y)$, is written as a functional of its trajectory $x(\tau)$ \cite{Majumdar2005,Sabhapandit2006}:
\begin{align}
\ell_t(y) = \int_0^t d \tau\, \delta(x(\tau)-y) \,. \label{lt}
\end{align}
Many properties of this quantity have been studied and analysed in various contexts over the years. Examples include diffusion in bounded domains \cite{Grebenkov2007}, diffusion in a potential landscape \cite{Majumdar2002}, diffusion in a random potential landscape \cite{Sabhapandit2006}, diffusion on a graph \cite{Comet2002}, time averaged statistical observables \cite{Lapolla2020}, single file problem \cite{Lapolla2018, Lapolla2019}, diffusion with reflected walls \cite{Forde2015, Fatlov2017}, and in conditioned diffusions \cite{Angeletti2016}. The local time has also been studied for random processes other than simple diffusion e.g., uniform empirical process \cite{Csorgo1999}, Brownian excursions \cite{Louchard1984} and for diffusion with stochastic resetting \cite{Pal2019}. In fact, more general functionals like 
\begin{align}
T_t[x(\tau)]= \int_0^t d \tau\, U(x(\tau)) \,, \label{T_t[x]}
\end{align}
 where $U(x)$ is an arbitrary function (with good properties), have been studied in different contexts by various methods~\cite{Majumdar2005,Grebenkov2007,Sabhapandit2006,Majumdar2002, Lapolla2020, Lapolla2018,Lapolla2019,Majumdar2002b,Touchette2018,Meerson2019,Nickelsen2018}.

In this paper, we consider an  over damped particle moving in one dimension in a viscous thermal medium, in the presence of an external potential $V(x)$.
The equation of motion for such a particle is
\begin{equation}
		\dot{x} = -V'(x) + \xi(t) \,, \label{eom}
\end{equation}
where $\xi(t)$ is a mean zero  Gaussian white noise with $\langle \xi(t)\xi(t')\rangle =2D \delta(t-t')$. Here, $\langle ... \rangle$ represents the noise realisation average, and $D$ controls the strength of the noise which would also be the diffusion constant of the particle.  We study the statistical properties of the local time $\ell_t(y)$ spent by the particle at some position $y$ in the presence of an external harmonic potential $V(x)=\kappa x^2/2$, with and without absorbing boundaries. For this choice of the potential $V(x)$, the motion in Eq.~\eqref{eom} is known in the literature as the Ornstein-Uhlenbeck (OU) process \cite{Gardiner1985,Risken1984}. 
In particular, mapping the problem to an equivalent quantum problem through the Feynman-Kac formalism~\cite{Kac1949,Kac1951}, we use the techniques of quantum perturbation theory to compute various moments of $\ell_t(y)$ as well as the equal time correlation $\langle \ell_t(y)\ell_t(z)\rangle$ among local times spent at different locations. This method is similar to the method described in~\cite{Grebenkov2007} where it was argued that within a bounded geometry, the cumulants of functionals like $T_t[x(\tau)]$ in Eq.~\eqref{T_t[x]} are all linear in $t$ for large $t$. This implies that the typical fluctuations of $T_t$ are of order $O(\sqrt{t})$ and are distributed according to a Gaussian. 

A similar perturbation method, based on the spectral theory of non-Hermitian operators, has recently been applied and discussed in more general contexts such as the study of  fluctuations in time-averaged statistical mechanics  \cite{Lapolla2020} and the single file problems \cite{Lapolla2018,Lapolla2019} where local time has been considered as a central object.  In \cite{Lapolla2020}, general expressions for the mean, variance, and covariance of an arbitrary functional, written in terms of eigenfunctions and matrix elements have been obtained, and from the  behaviour of these quantities at large time, the appearance of the Gaussian scaling distribution has been anticipated.

In this paper, applying quantum perturbation theory, we have obtained the mean, variance and covariance of the local time of an OU particle, expressed explicitly  in terms of Hermite polynomials in the presence and in the absence of an absorbing boundary. From the large $t$ behaviour of the  moments, we also find that the typical fluctuations are Gaussian in the presence of a confining potential, as found in \cite{Grebenkov2007,Lapolla2020}. 
However, in addition to the typical Gaussian distribution, we have computed the distribution of large deviations of the local time $\ell_t(y)$ as well. These are described by appropriate large deviation functions on the infinite line and on the semi infinite line with a reflecting boundary. 
For  reviews on Large deviation theory, see \cite{Touchette2018,Touchette2009}.

In the presence of absorbing boundaries, one usually studies statistics of path functionals such as $T_t[x(\tau)]$ (defined above) in two ways. The first way is to study the distribution of $T_t$ conditioned on survival over a given duration of time $t$, i.e.\@ over the ensemble of paths which survive till time $t$. In the second way, known as the study of first passage functionals, one is interested in the distribution of $T_t$ till the first passage time to the absorbing boundaries \cite{Majumdar2005,Majumdar2020}.  In this paper, we formulate a new general method to study these functionals for an arbitrary time duration $t$, not conditioned on survival, in the presence of a completely absorbing boundary. 
As it should, our formulation in the $t \to \infty$ limit corresponds to the study of first passage functionals.

The problem of path functionals in the presence of absorbing boundaries may appear in various situations --- for example, local time not conditioned on survival may appear in the context of reaction-diffusion processes inside a bounded domain with absorbing boundaries. Imagine catalytic reaction agents diffusing in a domain with heterogeneously distributed reaction centres. Once a particular agent reaches a centre, the corresponding reaction occurs. Such a catalytic reagent will continue to initiate reactions as long as it lives, i.e.\@ is not absorbed at the boundary. It is easy to understand that the total amount of production from reactions till time $t$ would depend on the time spent by reagents at these centres. In this paper, we present a general formulation for studying path functionals (defined in Eq.~\eqref{T_t[x]}) not conditioned on survival and apply this formalism to compute the moments  of $\ell_t(y)$ for an over damped particle evolving according to Eq.~\eqref{eom}.

We briefly summarise our results along with a presentation of the organisation of the paper. In Sec.~\ref{MGF}, we follow the Feynman-Kac formalism to describe how the computation of the the moment generating function (MGF) of the functional $T_t[x(\tau)]$, defined as $\mathcal{Q}(k,t|x_0) = \mean{e^{-kT_t[x(\tau)]}}_{x_0}$ with $x_0$ being the starting position, can be mapped to the problem of finding the eigenvalues and eigenfunctions of a quantum particle inside an appropriate confining potential. 
This is a well known mapping that has been observed and employed in various contexts to compute distributions of different observables \cite{Majumdar2005,Angeletti2016,Majumdar2002b,Touchette2018,Nyawo2018,Buisson2020,Meerson2019}.
The Hamiltonian of the quantum particle naturally appears to be a sum of two parts, in which the part involving $k$ can be treated as a perturbation to the other part. Using this structure, we apply quantum perturbation theory to compute the eigenvalues and eigenfunctions as a series expansion in powers of $k$, which is useful to directly compute various moments and correlations. 
As mentioned earlier, this perturbation method has been discussed and applied previously in different contexts \cite{Grebenkov2007, Lapolla2020}. 
To make this paper self-contained, we present the perturbation formalism for computing the moments of a general functional $T_t[x(\tau)]$ in Sec.~\ref{MGF}.
In Sec.~\ref{LT-harm-mean-variance}, we apply this formulation to the functional $\ell_t(y)$, defined in Eq.~\eqref{lt}, and obtain an exact, explicit expression for the mean $\mean{\ell_t(y)}$ (see Eq.~\eqref{<l_t(y)>-har-FS}) for arbitrary $t$, $y$ and initial position $x_0$. 
We have also obtained exact, explicit expressions for the asymptotic variance $\mean{\ell_t^2(y}_c$ and two-point correlation $\mean{\ell_t(y)\ell_t(z)}_c$ at large $t$ (see Eqs.~(\ref{<l_t^2(y)>-har-FS}) and (\ref{<l_t(y)l_t(z)>-har-FS})). 
We use $\mean{...}_c$ to represent cumulants and connected correlations. 

For cases in which the effective quantum confining potential is such that the ground state is non-degenerate and gapped, we find that at large time $t$, the cumulants of $T_t[x(\tau)]$ (defined in Eq.~\eqref{T_t[x]}) grow linearly with $t$. This suggests that the distribution of $T_t[x(\tau)]$, when shifted by its mean and scaled by its standard deviation, approaches a Gaussian distribution at large $t$. 
However, this distribution describes only the typical fluctuations, not the large fluctuations. In section \ref{LDF}, we describe how by performing a saddle point calculation on the leading term in the spectral representation of the MGF, one can establish the large deviation principle satisfied by a general functional $T_t[x(\tau)]$. 
This method has been used in other contexts  \cite{Majumdar2002b,Touchette2018, Nyawo2018, Buisson2020}. One usually relates the large deviation function to the dominant eigenvalue, i.e.\@ the ground state eigenvalue $\lambda_0(k)$ of the quantum problem through saddle point calculations, or equivalently, through a Legendre transform \cite{Touchette2009}. 
In Sec.~\ref{LDF-harm}, we apply this method to the local time $\ell_t(y)$, where we study analytic properties of the lowest eigenvalue $\lambda_0(k)$ in different regimes of $k$ and use these properties to find different asymptotic behaviours of the associated large deviation function for two choices of boundary conditions (see Eqs.~(\ref{LDF-asymp-harm-full}) and (\ref{LDF-reflect})). 
In the mathematics literature,  the large deviation function of functionals like $T_t[x(\tau)]$ (defined in Eq.~\eqref{T_t[x]})  can be computed from the large deviation theory developed by Donsker and Varadhan \cite{Touchette2009,Gartner1977}.
The large deviations for local time can in principle be obtained from their theory for the empirical distribution \cite{Touchette2009}. However, in this paper, we use a different method to compute exact, explicit expressions of large deviation function in different asymptotic limits. 

In the last part of the paper, starting in sec.~\ref{LT-NC-S}, we study statistical properties of the local time in the presence of an absorbing boundary, but not conditioned on survival. We compute the local time $\ell_t(y)$  spent by the OU particle at $y$ till time $t$. It is clear that if the particle survives, its trajectory may contribute to $\ell_t(y)$ i.e. may visit the location $y$ till time $t$. On the other hand, if it gets absorbed at time $t'<t$, then its trajectory will not contribute to $\ell_t(y)$ after time $t'$. However, we compute the local time spent at $y$ till time $t$ without the knowledge if the particle has already been absorbed at the boundary. It is easy to realise that in the $t \to \infty$ limit the distribution of $\ell_t(y)$ would approach a time independent (stationary) distribution $P_{\rm st}(\ell|x_0)$. We study this stationary distribution in  Sec.~\ref{sta-hp}; in the literature, this is known as the problem of first passage functionals \cite{Majumdar2005,Majumdar2020}. In this paper, we also study the approach to this stationary state 
by formulating a general method in Sec.~\ref{general-method} for computing the associated MGF at time $t$. We demonstrate this method by applying it to two cases: (i) Free particle ($V(x)=0$) and (ii) OU particle ($V(x)=\kappa x^2/2$).
For these two cases, we compute the approach of the mean local time to its stationary value with time (see Eqs.~(\ref{mean-lt-fp}) and (\ref{mean-lt-hp})).

\section{Moments and cumulants using Quantum perturbation theory}
\label{MGF}
To compute the statistics of the path functional $T_t[x(\tau)]$, we define the MGF as \cite{Majumdar2005}
\begin{equation}
	\mathcal{Q}(k,t|x_0) = \mean{\e^{-kT_t[x(\tau)]}}_{x_0} .
	\label{mcal_Q}
\end{equation}
The angle brackets in the above equation denote the expectation value over the distribution $P_{x_0}(T,t)$ of $T_t$, conditioned on survival of the particle which starts at $x_0$. Since the observable $T_t$ depends on the path as defined in Eq.~\eqref{T_t[x]}, this average can be transformed into an average over paths surviving till time $t$, whose probability density can be obtained from the probability density of the noise realisation $\xi(\tau)$, given by 
\begin{equation}
	\mathbb{P}(\l\{\xi(\tau)\r\}) =\frac{1}{\mathcal{N}}\op{exp}\l({-\frac{1}{4D}\int_0^t\xi^2(\tau)\d\tau}\r)
	\label{eq:P(xi(tau))}
\end{equation}
where $\mathcal{N}$ is the normalisation constant. The MGF $\mathcal{Q}(k,t|x_0)$ of $T_t[x(\tau)]$ can be written as 
\begin{equation}
\mathcal{Q}(k,t|x_0) = \mean{\e^{-k\int_0^tU(x(\tau))\d\tau}}_{x_0}=\frac{Q(k,t|x_0)}{Q(0,t|x_0)}
\label{mcal_Q_p}
\end{equation}
where, using the standard Ito calculus and the usual connections to the Schrodinger equation in quantum mechanics, one can show that 
\begin{align}
 Q(k,t|x_0) &= \int dx \int\displaylimits_{\mathclap{x(0)=z_0}}^{\mathclap{x(t)=x}} \mathcal{D}[x(\tau)]
 \operatorname{exp}\!\left(-\int_0^t\frac{d\tau}{4D} \left[ 
 \left(\frac{dx}{d\tau}+V'(x)\right)^2 -2D V''(x)+ 4 D k U(x)\right] \right)  \label{GF_1}\\
 &= \int dx \exp\!\left({-\frac{V(x)}{2D}}\right) \bra{x} \e^{-t\hat{H}_k} \ket{x_0} \exp\!\left({\frac{V(x_0)}{2D}}\right).\label{GF_2}
\end{align}
The operator $\hat{H}_k$ in the above equation is given by
\begin{equation}
	\hat{H}_k = -D\frac{d^2}{dx^2} + \mathcal{U}(x)+ kU(x)~~\text{with}~~
	\mathcal{U}(x) =\left( \frac{V'(x)^2}{4 D}-\frac{V''(x)}{2}\right), 
	\label{H_p}
\end{equation}
$V'(x)=d V(x)/dx$, and $V''(x)=d^2 V(x)/d x^2$ \cite{Majumdar2005,Touchette2018, Kurchan2009}.
Note that the quantity $Q(0,x_0,t)$ is actually the survival probability 
$S(t|x_0)$ of the particle in the domain $\mathcal{D}$ till time $t$, given that it started from $x(0)=x_0$. One has $Q(0,t|x_0)=1$ when there is no absorbing boundary in the system, i.e.\@ when the probability is conserved. Otherwise, in the presence of an absorbing boundary, $Q(0,t|x_0)<1$. 

The operator $\hat{H}_k$ in Eq.~\eqref{H_p} can be interpreted as the Hamiltonian of a quantum particle of mass $m =\frac{1}{2D}$, moving in an effective potential given by $\mathcal{U}(x)+kU(x)$ \cite{Majumdar2005}. 
For appropriate choices of $V(x)$ and $U(x)$ this potential can be a confining potential, and for such choices we expect the Hamiltonian $\hat{H}_k$ to have a discrete eigen spectrum.  
Let the eigenvalues and eigenfunctions of the `quantum' Hamiltonian $\hat{H}_k$, denoted in bra-ket notation by 
\begin{align}
\hat{H}_k \ket{\psi_{n,k}}=\lambda_n{(k)} \ket{\psi_{n,k}},~~~\text{where},~~\braket{x}{\psi_{n,k}} =\psi_{n,k}(x),~~
n=0,1,2, \ldots  \label{EV-eq}
\end{align}
be orthonormal eigenfunctions which satisfy the specified boundary conditions and $\lambda_{n+1}{(k)}>\lambda_n{(k)} \geq 0 ~\forall~n$.  We can then represent Eq.~\eref{GF_1} as
\begin{align}
	Q(k,t|x_0) =\int dx \exp\!\left({-\frac{V(x)}{2D}}\right) \left [ \sum_n   \e^{-\lambda_n{(k)}t} \psi_{n,k}(x)  \psi_{n,k}^*(x_0)\right] 
\exp\!\left({\frac{V(x_0)}{2D}}\right), \label{Q_U-def}
\end{align}
where $ \psi_{n,k}^*(x_0)$ is the complex conjugate of $\psi_{n,k}(x_0)$.
In this way the problem of calculating the moment generating function of a given statistic gets mapped to the problem of finding the eigenvalues and eigenstates of a quantum particle in the corresponding one dimensional potential, for which a variety of methods exist.

From the Taylor series expansion of the MGF $\mathcal{Q}(k,t|x_0)$, one can compute moments of $T_t$ at different order as 
\begin{equation}
	\mean{T_t^l} =   (-1)^l\frac{\d^l \mathcal{Q}(k,t|x_0)}{\d k^l} \bigg|_{k \to 0}= \frac{(-1)^l}{Q(0,t|x_0)} \frac{\d^l Q(k,t|x_0)}{\d k^l} \bigg|_{k \to 0} \,,\text{ for } l=1,2,3,\dots
	\label{moments_T}
\end{equation}
and the corresponding cumulants can be obtained from 
\begin{align}
\langle T_t^l \rangle_c =  (-1)^l \frac{\d^l \log \mathcal{Q}(k,t|x_0)}{\d k^l} \bigg|_{k \to 0}= (-1)^l\frac{\d^l}{\d k^l} \log \!\left(\frac{Q(k,t|x_0)}{Q(0,t|x_0)}\right) \bigg|_{k \to 0} \,, \text{ for } l=1,2,3, \dots
\label{cumulants_T}
\end{align}

In order to proceed, we note that the Hamiltonian $\hat{H}_k$ in \eref{H_p} is in the form $\hat{H}_k = \hat{H}_u + \hat{h}_k$ where $\hat{h}_k=kU(x)$. If the eigenvalues and eigenfunctions of $\hat{H}_u$ are known,
$\lambda_n{(k)}$ and $\psi_n{(k)}$ can be obtained considering $kU(x)$ as the perturbation Hamiltonian. According to the `quantum' perturbation theory, the perturbative expansions for the eigenstates and eigenvalues are (the superscript $(0)$ denotes the unperturbed quantities) \cite{Sakurai}
\begin{align}
	\lambda_n{(k)} ={}& \lambda_n^{(0)} + k \bra{\psi_n^{(0)}}U\ket{\psi_n^{(0)}} + k^2 \sum_{m\ne n}\frac{\abs{\bra{\psi_m^{(0)}}U\ket{\psi_n^{(0)}}}^2}{\lambda_n^{(0)}-\lambda_m^{(0)}} + \dots
	\label{lam_n-perturb} \\
	\begin{split}
		\ket{\psi_n{(k)}} ={}& \ket{\psi_n^{(0)}} + k \sum_{m\ne n} \frac{\bra{\psi_m^{(0)}}U\ket{\psi_n^{(0)}}}{\lambda_n^{(0)} - \lambda_m^{(0)}}\ket{\psi_m^{(0)}}  \\
		&+ k^2
		\custombracket{.}{2em}{[}
		\sum_{m\ne n}\sum_{p\ne n}\frac{\ket{\psi_m^{(0)}}\bra{\psi_m^{(0)}}U\ket{\psi_p^{(0)}}\bra{\psi_p^{(0)}}U\ket{\psi_n^{(0)}}}{\l(\lambda_n^{(0)}-\lambda_m^{(0)}\r)\l(\lambda_n^{(0)}-\lambda_p^{(0)}\r)}  \\
		 &\phantom{{} + k^2 \custombracket{.}{2em}{[} {} }
		 - \sum_{m\ne n}\frac{\ket{\psi_m^{(0)}}\bra{\psi_m^{(0)}}U\ket{\psi_n^{(0)}}\bra{\psi_n^{(0)}}U\ket{\psi_n^{(0)}}}{\l(\lambda_n^{(0)}-\lambda_m^{(0)}\r)^2} 
		\custombracket{.}{2em}{]}
		+ \dots
	\end{split}
	\label{psi_n-perturb}
\end{align}
We can substitute the above perturbative expansions of $\lambda_n{(k)}$ and $\ket{\psi_n{(k)}}$ in Eq.~\eqref{Q_U-def} and then in Eqs.~\eqref{moments_T} and \eqref{cumulants_T} to extract the required moments and cumulants of $T_t$. In the above expansions, we have assumed the eigenvalues are non-degenerate. For degenerate eigenvalues one needs to use appropriate perturbation theory \cite{Sakurai}.

The above method can be extended straightforwardly to compute correlations between multiple statistics (for example, the local times at two different points). One can replace the term $k U$ in ${\hat{h}_k}$ by $k U_1 + k' U_2$ and then follow the same procedure.
Extracting the coefficients of appropriate powers of $k$ and $k'$ from the MGF would provide the required moments up to a multiplicative constant.
In fact, one can consider a vector of statistics defined by ${\bf T}[x(\tau)] \defn \inSet{T_i[x(\tau)]}=\inSet{T_1[x(\tau)],T_2[x(\tau)],\dots,T_N[x(\tau)]}$, along with a real vector ${\bf k} \defn \inSet{k_1,k_2,\dots,k_N}$ of the same dimension $N$.
An extended moment generating function can be defined as $Q({\bf k},t|x_0) \defn\mean{e^{-{\bf k}\cdot {\bf T}[x(\tau)]}}$, from which one can get
\begin{align}
	\mean{T_{1}^{l_1}\dots T_{r}^{l_r}} &= \frac{(-1)^{l_1+l_2+\dots+l_r}}{S(t|x_0)}\evalAt{ \frac{\dh^{l_1}}{\dh k_{1}^{l_1}}\dots \frac{\dh^{l_r}}{\dh k_{r}^{l_r}} Q(k_1,k_2,\dots,k_R,t|x_0) }_{\vec{k}=0} \label{multi-T-corr}\\
\mean{T_{1}^{l_1}\dots T_{r}^{l_r}}_c &= 	(-1)^{l_1+l_2+\dots+l_r}\evalAt{ \frac{\dh^{l_1}}{\dh k_{1}^{l_1}}\dots \frac{\dh^{l_r}}{\dh k_{r}^{l_r}} \log \left [\frac{Q(k_1,k_2,\dots,k_R,t|x_0)}{Q(0,t|x_0)} \right] }_{\vec{k}=0}
\label{multi-T-corr-con}
\end{align}
where $l_i=0,1,2,\dots$ with $\sum_{i=1}^rl_i \neq 0$  and $1\leq r\leq N$. In Sec.~\ref{LT-harm}, we use the above-mentioned perturbative approach to obtain explicit expressions for the mean, variance and covariance of the local time spent by the OU particle at different locations.

\section[{Distribution of Tt[x(tau)]: Typical and atypical fluctuations}]{Distribution of $T_t[x(\tau)]$: Typical and atypical fluctuations}
\label{LDF}
While the perturbative approach allows us to evaluate a few lower order moments as well as cumulants of an arbitrary functional $T_t[x(t)]$ (of the form in Eq.~\eqref{T_t[x]}) in an external potential, the expressions obtained involve nested summations which may be difficult to evaluate. Note that to obtain an exact explicit expression for the $l^{th}$ moment, we need $l^{th}$ order perturbation theory. This approach is feasible for lower-order moments (such as mean, variance), but quickly becomes tedious and cumbersome for higher-order moments. However, for large $t$, it is easy to see that the leading term of cumulants of all order grows linearly with time, as follows: We first write the MGF $Q(k,t|x_0)$ in Eq.~\eqref{Q_U-def}  as
\begin{align}
\begin{split}
Q(k,t|x_0)&=\sum_n   \e^{-\lambda_n{(k)}t}g_n(k) f_n(k,x_0) \\
&= \e^{-\lambda_0 {(k)}t} \left[ g_0(k)f_{0}(k,x_0) +   \e^{-(\lambda_1{(k)}-\lambda_0 {(k)}) t} g_1(k)f_{1}(k,x_0)+ \dots \right]
\end{split} \label{Q(k)-expand}
\end{align}
where
\begin{align}
\begin{split}
f_{n}(k,x_0) &= \psi_{n,k}(x_0)
\exp\!\left({\frac{V(x_0)}{2D}}\right) \\ 
g_{n}(k) &= \int dx \exp\!\left({-\frac{V(x)}{2D}}\right) \psi_{n,k}(x) \,.
\end{split}
\label{mcalQ}
\end{align}
In the above, we have assumed that the eigenvalues inside a confining potential satisfy $0\leq \lambda_0  < \lambda_1 \leq \lambda_2 \dots$ even in the presence of perturbation.
Using this expression in Eq.~\eqref{cumulants_T} and taking the large $t$ limit, we get 
\begin{align}
\langle T_t^l \rangle_c &=  (-1)^l \frac{\d^l \log [Q(k,t|x_0)/Q(0,t|x_0)]}{\d k^l} \bigg|_{k \to 0} \nonumber \\ 
&\simeq  t \left[(-1)^{l+1}\frac{\d^l \lambda_0 {(k)}}{\d k^l} \right]_{k \to 0} + ~\text{constant~term}~+~\text{terms~exponentially~small~in}~t
\end{align}
for $l=1,2,3,\dots$. Hence, we observe that at large $t$, all cumulants in the leading order increase linearly. Consequently, the distribution of the scaled random variable $q={(T_t - \langle T_t \rangle)}/{\sqrt{\langle T_t^2 \rangle_c}}$ approaches a Gaussian distribution i.e.\@  for large $t$ we have
\begin{align}
P_{x_0}(T_t=T,t) \simeq \frac{1}{\sqrt{\langle T_t^2 \rangle_c}}\op{\mathbb{G}}\!\left( \frac{T - \langle T_t \rangle}{\sqrt{\langle T_t^2 \rangle_c}}\right) ,~~~
\text{with}~~\mathbb{G}(q)=\frac{1}{\sqrt{2 \pi}} \exp\!\left( -\frac{q^2}{2}\right),
\label{T-typ-G}
\end{align}
as obtained in \cite{Grebenkov2007,Lapolla2020}.
The distribution in the above equation describes typical fluctuations of $T_t$ of order $O(\sqrt{t})$ around $\mean{T_t}$, but it  does not describe the probabilities of rare fluctuations such as very large or small values of $T_t$ compared to $\mean{T_t}$. 
For this, we need to study the large deviation properties of $T_t$, which we present later. The result in Eq.~\ref{T-typ-G}  is valid under the assumption that the the eigenspectrum of the quantum Hamiltonian $\hat{H}_u$ has a gap above the non-degenerate ground state eigenvalue $\lambda_0^{(0)}$, and that this remains true even after adding the perturbation $k U(x)$. Under these assumptions, the result in Eq.~\ref{T-typ-G} can be easily generalised to the joint distribution of multiple observables ${\bf T}_t[x(\tau)] = \{T_1[x(\tau)],T_2[x(\tau)],\dots,T_N[x(\tau)]\}$. In the large $t$ limit, one can show that the joint distribution of the normalised observables ${\bm \zeta}=\{\zeta_i\}=\{~(T_i - \langle T_i \rangle)/\sqrt{\langle T_i^2 \rangle_c}~\}$, obtained after proper shifting and scaling, is a multivariate Gaussian.

We now proceed to describe a non-perturbative method of finding  the large-deviation functions of the functional $T_t[x(\tau)]$, defined in Eq.~\eqref{T_t[x]}. As mentioned earlier, in the large $t$ limit, the $0$th eigenvalue $\lambda_0 (k) $ contributes at the leading order. 
Hence, we can write 
\begin{align}
Q(k,t|x_0)\simeq  \e^{-\lambda_0 (k)t} g_0(k)f_0(k,{x_0})
\label{mcalQ-aprx}
\end{align}
for large $t$, where $g_0(k)$ and $f_0(k,x_0)$ are given in Eq.~\eqref{mcalQ}. Observe that $\mathcal{Q}(k,t|x_0)$ in Eq.~\eqref{mcal_Q_p} is basically the Laplace transform of $P_{x_0}(T,t)$ with respect to $T$. Hence, performing the inverse Laplace transform of $\mathcal{Q}(k,t|x_0)$, one finds that 
\begin{align}
P_{x_0}(T,t) \approx \frac{1}{S(t|x_0)} \int dk \exp\left(-t~\left[-k~\frac{T}{t}  +\lambda_0 {(k)} \right]~\right) g_0(k)f_0(k,{x_0}) 
\label{ILT-mcalQ_p}
\end{align}
for large $t$. 
In the above, $S(t|x_0)$ is the survival probability, given by
\begin{equation}
	S(t|x_0) = Q(0,t|x_0) = \sum_n \e^{-t\lambda_n^{(0)}}g_n^{(0)}f_n^{(0)}(x_0) \,. \label{S(t|x_0)}
\end{equation}
The integral in Eq.~\eqref{ILT-mcalQ_p} can be performed using the steepest-descent method \cite{Arfken}. 
In order to  do that, one needs to know $\lambda_0 (k)$ 
for all $k$. The expression in Eq.~\eqref{lam_n-perturb} provides $\lambda_0 (k)$ for small $k>0$.
While solving the eigenvalue problem in Eq.~\eqref{EV-eq}, one first writes the general solutions and then tries to fix the integration constants by satisfying the boundary conditions. Often, demanding non-zero and non-trivial values for the integration constants provides an equation involving $\lambda_0 $ and $k$, 
solving which, one obtains the desired eigenvalues for each value of $k$. For example, in the case of the local time density $\ell_t(y)$ (in which we are mainly interested in this paper), this equation, for different boundary conditions, has the form 
\begin{align}
G(\lambda_0 , y) =k \,. \label{eg-sol-eq}
\end{align}
We will observe this fact in the next section, where we compute this relation explicitly for a harmonic confining potential with different boundary conditions.

In the steepest-descent method, the dominant contribution comes from the saddle point $k^*$ where the argument of the exponential in Eq.~\eqref{ILT-mcalQ_p} is maximum, i.e.\@
\begin{align}
\left[\frac{d \lambda_0 (k)}{d k}\right]_{k=k^*}=w,~~\text{where},~~w=\frac{T}{t} \,.
\end{align}
Now, from Eq.~\eqref{eg-sol-eq}, one finds $d \lambda_0  /dk= [\partial_{\lambda_0 }G(\lambda_0 ,y)]^{-1}$. Hence, $\lambda_0 ^*(w,y)=\lambda_0 (k^*(w,y))$ is obtained by solving
\begin{equation}
[\partial_{\lambda_0 }G(\lambda_0 ,y)]_{\lambda_0 =\lambda_0 ^*}=w^{-1} \,, \label{lam-star}
\end{equation}
which is then used to perform the saddle point integration in Eq.~\eqref{ILT-mcalQ_p} to get  
\begin{align}
P_{x_0}(T,t)& \underset{\text{large}~t}{ \simeq } \frac{\mathcal{P}_{x_0}(T/t,~t)}{S(t|x_0)},~~~\text{where} \\
\mathcal{P}_{x_0}(w,t)&= 
\frac{\exp\!\left\{ -t\left[ -w G(\lambda_0 ^*,y)+\lambda_0 ^*\right]\right\} g_0(G(\lambda_0 ^*,y)) \,f_{0}(G(\lambda_0 ^*,y),x_0)}{\sqrt{2 \pi t~\Big{|}\left(\frac{\partial_{\lambda}^2G(\lambda,y)}{(\partial_\lambda G(\lambda,y))^3}\right)_{\lambda=\lambda_0 ^*} \Big{|} }} \,. 
\label{P_U_t}
\end{align}
This is a general formula for the distribution of the local time density $\ell_t$ (defined in Eq.~\eqref{lt}) spent at $y$ by an overdamped particle moving inside a confining potential $V(x)$.

The above discussion on saddle point computation, presented in the context of local time, can be straightforwardly generalised to other functionals of the form $T_t[x(\tau)]$ (in Eq.~\eqref{T_t[x]}). For such functionals, the equivalent of equation \eqref{eg-sol-eq} would be of different form, but would still provide a relation between $\lambda_0 $ and $k$. For small $k$, this relation would provide an expansion of the form in Eq.~\eqref{lam_n-perturb}. However, for arbitrary $k$, one would be able to find $\lambda_0 $ numerically or through asymptotic expansions. Such evaluations can then be used to find the 
saddle point for a given value of $T_t[x(\tau)]$ and $t$, which would finally provide a large deviation form similar to Eq.~\eqref{P_U_t}.

In the next section, we explicitly calculate $\ell_t(y)$ for an OU particle with different boundary conditions.

\section{Local time statistics in a harmonic potential}
\label{LT-harm}
\subsection{Mean, variance and covariance}
\label{LT-harm-mean-variance}
In this section, we apply the methods discussed and presented in sections \ref{MGF} and \ref{LDF} to study statistical properties of the local time functional (defined in Eq.~\eqref{lt}) for an OU particle, i.e.\@ for $V(x)=\kappa  x^2/2,~\kappa >0$. For this choice, the Hamiltonian $\hat{H}_k$ in Eq.~\eqref{H_p} becomes 
\begin{align}
\hat{H}_k=\hat{H}_u + \hat{h}_k,~\text{where}~\hat{H}_u=-D\frac{\dh^2}{\dh x^2}+\frac{\kappa ^2}{4D}x^2-\frac{\kappa }{2}
~~\text{and}~~~\hat{h}_k=k\,\delta(x-y) \,.
\label{H-harm-loc}
 \end{align}
The delta function in the above appears from the definition of  local time $\ell_t(y)$ in Eq.~\eqref{lt}. To compute the moments and cumulants, we will compute the characteristic function $Q(k,t|x_0)$  defined in Eq.~\eqref{GF_2}, explicitly as a series in powers of $k$ using perturbation theory. Before going into that, we point out that in this case, because of the particular delta function form of the perturbation Hamiltonian $\hat{h}_k$ the Laplace transform of the distribution $P_{x_0}(\ell,t)$ of $\ell$ with respect to $t$ can be expressed in closed form
 in terms of the Green's function of the unperturbed Hamiltonian
 \cite{comtet-HP}. 

However, in this paper we apply perturbation theory in time domain explicitly, as it seems to be easily generalisable to study of local times at multiple locations. For example, if we are interested in local times at positions $y$ and $z$, then we should consider the perturbation Hamiltonian  $\hat{h}_{\bf k}=k\delta(x-y)+k'\delta(x-z)$. Similarly, to study local times at multiple locations one needs to modify the perturbation Hamiltonian accordingly.  

Note that $\hat{H}_u$ is the Hamiltonian of a quantum harmonic oscillator, for which the eigenvalues and eigenfunctions can be obtained for specified boundary conditions. Once they are known, one can compute the eigenvalues and eigenfunctions of $\hat{H}_{\bf k}$ by using the perturbative expansions in Eqs.~\eqref{lam_n-perturb} and \eqref{psi_n-perturb}. Substituting such expansions in the expression of $Q({\bf k},t|x_0)$ (Eq.~\eqref{Q(k)-expand}), one can also expand it in powers of $k_i$'s from which one would be able to find correlations and moments/cumulants from Eqs.~\eqref{multi-T-corr}. For example, to compute the  correlation between local times at two locations, $y$ and $z$, we have the following perturbation expansion of the MGF $Q({\bf k},t|x_0)$ (Eq.~\eqref{Q(k)-expand}):
\begin{align}
\begin{split}
&Q({\bf k},t|x_0) = \sum_n  \e^{-t( \lambda_n^{(0)}+k \lambda_n^{(1)}(y)+k' \lambda_n^{(1)}(z)+k^2  \lambda_n^{(2)}(y)+k'^2  \lambda_n^{(2)}(z)+kk'\lambda_n^{(1,1)}(y,z)+\dots)} \\
&\quad \times \left[ g_n^{(0)}+kg_n^{(1)}(y)+k'g_n^{(1)}(z)+k^2g_n^{(2)}(y)+k'^2g_n^{(2)}(z)+kk'g_n^{(1,1)}(y,z)+\dots\right]
\label{Q_k_expand} \\
&\quad \times \left[ f_n^{(0)}(x_0)+kf_n^{(1)}(y|x_0)+k'f_n^{(1)}(z|x_0)+k^2f_n^{(2)}(y|x_0)+k'^2f_n^{(2)}(z|x_0) \right. \\
& \phantom{mmmmmmmmmmmmmmmmmmmmmmmmmm} \left. +kk'f_n^{(1,1)}(y,z|x_0)+\dots\right] .
\end{split}
\end{align}
where expressions for terms like $\lambda_n^{(0)},~\lambda_n^{(1)},\dots$; $g_n^{(0)},~g_n^{(1)},\dots$; and \sloppy$f_n^{(0)}(x_0),~f_n^{(1)}(y|x_0),\dots$ can be obtained from Eqs.~\eqref{lam_n-perturb} and \eqref{psi_n-perturb} by replacing $kU$ by $\hat{h}_{\bf k}=k\delta(x-y)+k'\delta(x-z)$.  The explicit expressions are provided in \ref{higher-order-terms}. Substituting the expressions from Eqs.~(\ref{lam_n^1}--\ref{g_n^11}) in Eq.~\eqref{Q_k_expand}, we get the expansion of $Q({\bf k},t|x_0)$ in powers of $k$ and $k'$ up to second order. Inserting this expansion in Eq.~\eqref{multi-T-corr}, one gets the first two moments and two-point correlations as 
\begin{align}
\mean{\ell_t(y)} &=  t~\frac{\sum_n \e^{-t\lambda_n^{(0)}}\lambda_n^{(1)}(y)g_n^{(0)}f_n^{(0)}(x_0)}{S(t|x_0)} 
- \frac{\sum_n \e^{-t\lambda_n^{(0)}}[g_n^{(1)}(y)f_n^{(0)}(x_0)+g_n^{(0)}f_n^{(1)}(y|x_0)]}{S(t|x_0)} \,, 
\label{<l_t(y)>}\\ 
\mean{\ell_t^2(y)}&=t^2~\frac{\sum_n   \e^{-t\lambda_n^{(0)}}\lambda_n^{(1)}(y)^2g_n^{(0)}f_n^{(0)}(x_0)}{S(t|x_0)} \nonumber \\
&~- 2t~\frac{\sum_n   \e^{-t\lambda_n^{(0)}}\left[\lambda_n^{(2)}(y)g_n^{(0)}f_n^{(0)}(x_0) +\lambda_n^{(1)}(y)\left(g_n^{(0)}f_n^{(1)}(y|x_0)+g_n^{(1)}(y)f_n^{(0)}(x_0)\right) \right] }{S(t|x_0)} \nonumber \\
&~+ { 2}\frac{\sum_n   \e^{-t\lambda_n^{(0)}}\left[g_n^{(2)}(y)f_n^{(0)}(x_0) + g_n^{(0)}f_n^{(2)}(y|x_0) +g_n^{(1)}(y)f_n^{(1)}(y|x_0) \right] }{S(t|x_0)} \,,~~~\text{and} 
\label{<l_t^2(y)>}
\end{align} 
\begin{equation}
\resizebox{\textwidth}{!}{$
\begin{aligned}
&\langle \ell_t(y)\ell_t(z)\rangle =
{  t^2 \, \frac{ \sum_n \e^{-t\lambda_n^{(0)}} f_n^{(0)}(x_0) g_n^{(0)} \lambda_n^{(1)}(y) \lambda_n^{(1)}(z) }{S(t|x_0)} }
- t \left[\frac{\sum_n   \e^{-t\lambda_n^{(0)}}{ \lambda_n^{(1,1)}(y,z)}g_n^{(0)}f_n^{(0)}(x_0) }{S(t|x_0)} \right. \\ 
&~+ \left.\frac{\sum_n   \e^{-t\lambda_n^{(0)}}\left[ f_n^{(0)}(x_0)\left(g_n^{(1)}(y)\lambda_n^{(1)}(z) +g_n^{(1)}(z)\lambda_n^{(1)}(y)\right) +  g_n^{(0)}\left(f_n^{(1)}(y|x_0)\lambda_n^{(1)}(z) +f_n^{(1)}(z|x_0)\lambda_n^{(1)}(y)\right) \right] }{S(t|x_0)} \right] \\
&~+\frac{\sum_n   \e^{-t\lambda_n^{(0)}}\left[g_n^{(1,1)}(y,z)f_n^{(0)}(x_0)+f_n^{(1,1)}(y,z|x_0)g_n^{(0)}+g_n^{(1)}(y)f_n^{(1)}(z|x_0)+f_n^{(1)}(y|x_0)g_n^{(1)}(z)\right] }{S(t|x_0)} \,.
\label{<l_t(y)l_t(z)>}
\end{aligned}
$}
\end{equation}
where $S(t|x_0)$ is the survival probability, defined in Eq.~\eqref{S(t|x_0)}.
From these expressions, one can get the cumulants as $\mean{\ell_t(y)^2}_c= \mean{\ell_t(y)^2}-\mean{\ell_t(y)}^2$ and $\mean{\ell_t(y)\ell_t(z)}_c= \mean{\ell_t(y)\ell_t(z)}-\mean{\ell_t(y)}\mean{\ell_t(z)}$. 
One could also compute these quantities directly by substituting the expression of $Q({\bf k},t|x_0)$ from Eq.~\eqref{Q_k_expand} in Eq.~\eqref{multi-T-corr-con}. 
The above general expressions for the mean, variance, and covariance of the local time of an OU particle are similar to those obtained in \cite{Lapolla2020}.
In this paper, we provide  {\it more explicit} expressions for these cumulants in the presence and in the absence of an absorbing boundary.

\subsubsection{On infinite line:}
\label{LT-line}
In this case, the eigenvalues and eigenfunctions of the bare Hamiltonian $\hat{H}_u$ with the boundary conditions $\psi_n(x) \to 0$ as $x \to \pm \infty$ are given by  \cite{Griffiths2005,Shankar1994}
 \begin{align}
 \begin{split}
 \lambda_n^{(0)} &= n \kappa \,,~~~\text{and} \\
 	\psi^{(0)}_n(x) &=\braket{x}{\psi_n^{(0)}}= \frac{1}{\sqrt{2^nn!}}\l(\frac{\kappa }{2\pi D}\r)^{\frac{1}{4}} \e^{-\frac{\kappa x^2}{4D}}{H}_n\!\left(\sqrt{\frac{\kappa }{2D}}x\right) \,,
	\end{split}
	\label{psi_n^0(x)-IL}
 \end{align}
 for $n=0,1,2,\dots$ where ${H}_n(z)$ is the Hermite polynomial of $n$th degree. Note that in this case, the survival probability 
 $S(t|x_0)= \sum_ne^{-t\lambda_n^{(0)}}g_n^{(0)}f_n^{(0)}(x_0)=1$ for any $x_0$, because $g_n^{(0)} =\delta_{n,0},~\lambda_0^{(0)}=0$ and $f_0^{(0)}(x_0) =1$.

\begin{figure}
	\centering
		\includegraphics[width=0.32\textwidth]{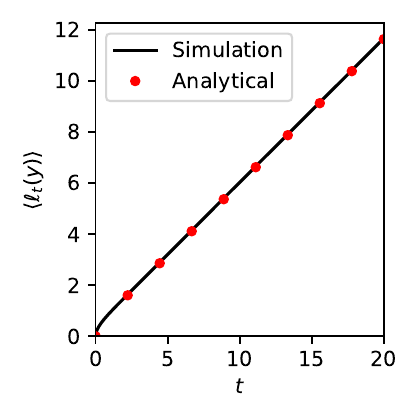}%
		\includegraphics[width=0.32\textwidth]{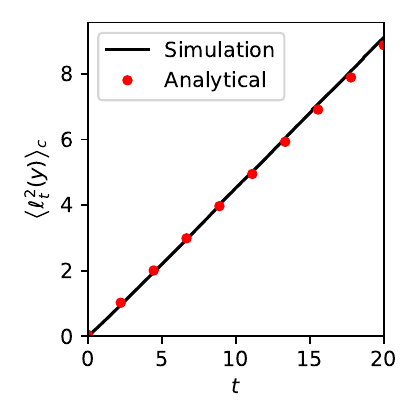}%
		\includegraphics[width=0.35\textwidth]{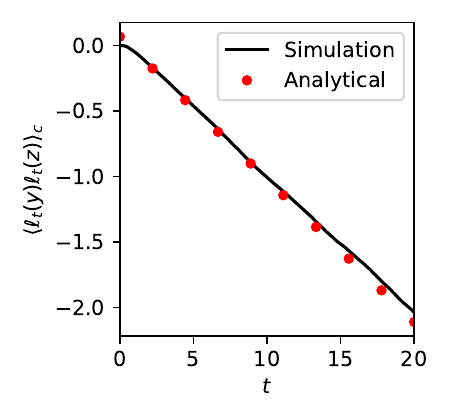}
	\caption{Comparison of numerical and analytical results for the mean local time $\mean{\ell_t(y)}$ (left) in Eq.~\eqref{<l_t(y)>-har-FS}, the variance  $\mean{\ell_t^2(y)}_c$ (middle) in Eq.~\eqref{<l_t^2(y)>-har-FS} and the covariance $\mean{\ell_t(y)\ell_t(z)}_c$ (right) in Eq.~\eqref{<l_t(y)l_t(z)>-har-FS} for $y=0$, $z=1$ with $x_0=0$.  We have chosen $D=1/2$ and $\kappa =1$. For the simulations, we used a forward Euler scheme with time-step $dt=10^{-5}$, and took $10^5$ realizations.}
	\label{fig:<l_t(y)>-har-FS}
\end{figure}

 Substituting the above  eigenvalues and eigenfunctions in Eqs.~(\ref{lam_n^1}--\ref{g_n^11}), we first find the functions $\lambda_n^{(..)}$, $g_n^{(..)}$ and $f_n^{(..)}$. These functions are then substituted in Eq.~\eqref{<l_t(y)>}, Eq.~\eqref{<l_t^2(y)>}, and Eq.~\eqref{<l_t(y)l_t(z)>} to find the first and second moments and the two point correlation $\langle \ell_t(y) \ell_t(z) \rangle$. In particular, we find the following exact, explicit expression for the mean:
 \begin{align}
\langle \ell_t (y)\rangle = \sqrt{\frac{\kappa }{2 \pi D}}  \e^{-\frac{\kappa  y^2}{2D}} \left(t + \sum_{m=1}^\infty \frac{(1-\e^{-m\kappa t})}{m\kappa }  \frac{H_m\!\left(\sqrt{\frac{\kappa }{2D}}y\right)H_m\!\left(\sqrt{\frac{\kappa }{2D}}x_0\right)}{2^m m!}\right),
\label{<l_t(y)>-har-FS}
\end{align}
where $x_0$ is the point at which the OU particle starts its motion, and the local time is measured at the point $y$.
 This expression is numerically verified in the leftmost panel of Fig.~\ref{fig:<l_t(y)>-har-FS}.
Note from Eq.~\eqref{<l_t(y)>-har-FS} that the average of the fraction of local time, $\langle \ell/t \rangle$, converges in the large $t$ limit to the stationary position distribution of the OU particle.
This is expected, since the particle dynamics is ergodic \cite{Bel2005occ,Barkai2006}.

The second cumulant and the two point correlations can also be evaluated from Eq.~\eqref{<l_t^2(y)>} and Eq.~\eqref{<l_t(y)l_t(z)>}.
Here, we do not present their expressions, as they are lengthy and involve multiple infinite series.
Instead, we focus on the large $t$ limit, where one can make further simplifications. Moreover, we assume $D=1/2$, $y=0$ and $x_0=0$ to simplify the expression for the variance further. In the large $t$ limit, we find
\begin{align}
	\mean{\ell_t(0)^2}_c \underset{t >> \frac{2\pi}{\kappa }}{ \simeq }& 
		 \frac{1}{2D}\left[\frac{2\log2}{\pi}t
		+ \frac{\mathcal{A}}{\kappa}
		\right] 
		, \text{ where } \mathcal{A} \approx 0.044
		\label{<l_t^2(y)>-har-FS} \\
\mean{\ell_t(y)\ell_t(z)}_c \underset{t >> \frac{2\pi}{\kappa }}{ \simeq }&	\left[  \frac{1}{\pi D} \e^{-\frac{\kappa y^2}{2D}-\frac{\kappa z^2}{2D}}
\sum_{m=1}^\infty
\frac{H_m\!\left(\sqrt{\frac{\kappa}{2D} }y\right)H_m\!\left(\sqrt{\frac{\kappa}{2D} }z\right)}{m!\,m2^m} \right] t +\mathcal{C}_{\langle \ell \ell \rangle }(y,z|x_0) \,, 	\label{<l_t(y)l_t(z)>-har-FS} 
\end{align}
where $\mathcal{C}_{\langle \ell \ell \rangle }(y,z|x_0)$ is given in Eq.~\eqref{C_(ll)}. 
In the middle panel of fig.~\ref{fig:<l_t(y)>-har-FS}, we compare Eq.~\eqref{<l_t^2(y)>-har-FS} and in  the right panel we compare Eq.~\eqref{<l_t(y)l_t(z)>-har-FS} 
with simulations; we observe agreement in both cases.
\subsubsection{On semi-infinite line: Local time statistics conditioned on survival -- }
\label{LT-C-S}
\begin{figure}
	\centering
	\includegraphics[width=0.33\textwidth,keepaspectratio]{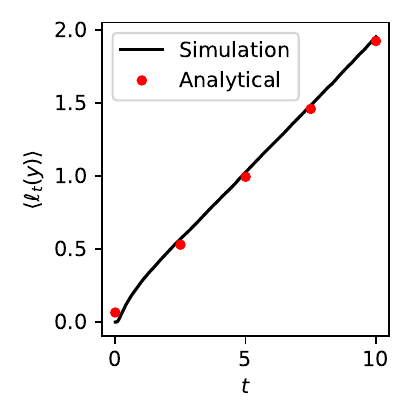}%
	\includegraphics[width=0.33\textwidth,keepaspectratio]{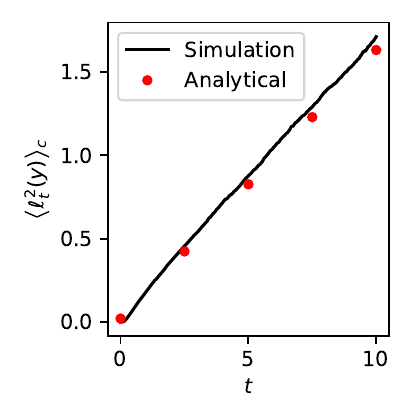}%
	\includegraphics[width=0.33\textwidth,keepaspectratio]{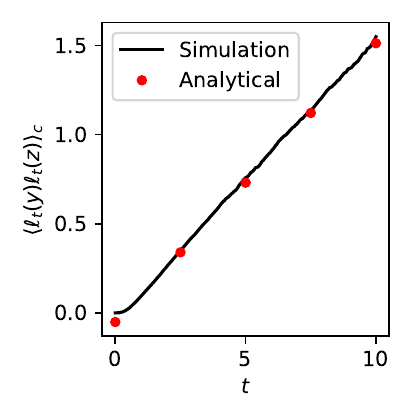}%
	\caption{ The analytical expressions for the mean, the variance, and the two point correlation given in Eqs.~\eqref{<l>_a} and \eqref{<ll>_a} are verified by comparing them with numerical simulations. In the left panel, we plot the mean $\mean{\ell_t(y)}$. In the middle panel, we plot the variance $\mean{\ell_t(y)}_c$, and in the right panel, we plot the covariance $\mean{\ell_t(y)\ell_t(z)}_c$ for $y=0.3$ and $z=0.6$. The other parameters of this plot are: 
	$x_0=0.9$, $\kappa=1$, and $D=1/2$. For the simulations, we used a forward Euler scheme with time-step $dt=10^{-5}$. Averages were taken over $\sim 10^5$ surviving realizations.} 
	\label{fig:Num-abs-bc}
\end{figure}

In this case, we consider a particle starting from some position $x_0>0$, moving on the positive semi-axis, with an absorbing boundary at the origin ($x=0$). The eigenfunctions of the bare Hamiltonian $\hat{H}_u$ need to satisfy the boundary conditions $\psi_n(x) \to 0$ as $x \to  \infty$ and $\psi_n(x=0)=0$.  The eigenvalues and eigenfunctions of the bare Hamiltonian $\hat{H}_u$ are then given by \cite{Griffiths2005,Shankar1994}
\begin{align}
	\left.\begin{aligned}
		\lambda_n^{(0)} &= n \kappa \\
		\psi^{(0)}_n(x) &=\braket{x}{\psi_n^{(0)}}= \frac{\sqrt{2}}{\sqrt{2^nn!}}\l(\frac{\kappa }{2\pi D}\r)^{\frac{1}{4}} \e^{-\frac{\kappa x^2}{4D}}\hat{H}_n\!\left(\sqrt{\frac{\kappa }{2D}}x\right)
	\end{aligned}\right\}
	\text{ for $n=1,3,5,\dots$}
\label{psi_n^0(x)-IL-ab}
\end{align}
Note that in this case, the survival probability $S(t|x_0)=Q(0,t|x_0) \neq 1$ for $x_0 > 0$.
In fact, we have
\begin{align}
S(t|x_0)= \sum_{n=1,3,5\dots}e^{-t\lambda_n^{(0)}}&g_n^{(0)}f_n^{(0)}(x_0) = \sum_{n=1,3,5\dots} \sqrt{\frac{2}{2^n n! }} \,\mathcal{J}_{0,n} \,H_n\!\left(\sqrt{\frac{\kappa }{2D}}x_0\right) \e^{-n\kappa t},\\ 
&\text{where},~
	\mathcal{J}_{0,n}=  \sqrt{\frac{2}{2^n n! \pi}}\int_0^\infty du\, \e^{-u^2} H_n(u) \,.
\end{align}
As in Sec.~\ref{LT-line}, we substitute the above  eigenvalues and eigenfunctions in Eqs.~(\ref{lam_n^1}--\ref{g_n^11}). We find the functions $\lambda_n^{(..)}$, $g_n^{(..)}$ and $f_n^{(..)}$, and then substitute them in Eq.~\eqref{<l_t(y)>}, Eq.~\eqref{<l_t^2(y)>}, and Eq.~\eqref{<l_t(y)l_t(z)>} to find the first and second moments and the two point correlation $\langle \ell_t(y) \ell_t(z) \rangle$. In particular, for large $t$ we find the following explicit asymptotic expressions:
\begin{align}
		\mean{\ell_t(y)} \underset{t >> \frac{2\pi}{\kappa }}{ \simeq } {}&  \sqrt{\frac{\kappa}{2 \pi D}}
		\frac{2 \kappa \, y^2 }{D} \e^{-\frac{\kappa y^2}{2D}} t
		+ \frac{1}{\sqrt{\pi}} 
		\frac{y}{D} \e^{-\frac{\kappa y^2}{2D}}
		\sum_{\substack{m\ne1\\\text{$m$ odd}}} \frac{(-1)^\frac{3m-1}{2} m!!\, H_m\!\left(\sqrt{\frac{\kappa}{2D}}y\right)}{2^\frac{m+1}{2} m! \l(1-m\r) m} \nonumber \\
		& - \frac{2}{\sqrt{2 \kappa \pi D}}
		\frac{y}{x_0} \e^{-\frac{\kappa y^2}{2D}}
		\sum_{\substack{m\ne1\\\text{$m$ odd}}} \frac{H_m\!\left(\sqrt{\frac{\kappa}{2D}}y\right)H_m\!\left(\sqrt{\frac{\kappa}{2D}}x_0\right)}{2^m m! \l(1-m\r)} \label{<l>_a} \\
		&+\text{terms~decaying~exponentially~in~}t  \nonumber 
\\
\begin{split}
\mean{\ell_t(y)\ell_t(z)}_c \underset{t >> \frac{2\pi}{\kappa }}{ \simeq } {}& 
- \frac{4 \kappa \, y z }{\pi D^2}  
		\e^{-\frac{\kappa y^2}{2D}-\frac{\kappa z^2}{2D}}
		\sum_{\substack{m\ne1\\\text{$m$ odd}}} \frac{H_m\!\left(\sqrt{\frac{\kappa}{2D}}y\right)H_m\!\left(\sqrt{\frac{\kappa}{2D}}z\right)}{2^m m! \l(1-m\r)} ~t  \\
		& +  \mathcal{C}_{\langle \ell \ell \rangle}^{(a)}(y,z|x_0)
		+ \text{terms~decaying~exponentially~in~}t
\end{split}
\label{<ll>_a}
\end{align} 
where $\mathcal{C}_{\langle \ell\ell \rangle}^{(a)}(y,z|x_0)$ is given in Eq.~\eqref{C_(ll)^a}. The variance $\mean{\ell_t^2(y)}_c$ can be obtained from Eq.~\eqref{<ll>_a} by putting $z=y$. These expressions for the mean, variance, and covariance are verified in fig.~\ref{fig:Num-abs-bc} where we observe nice agreement.  

\subsection{Typical fluctuations and large deviations}
\label{LDF-harm}
To compute the probability of large deviations of local time $\ell_t(y)$ at a location $y$, we would apply the formalism 
described in section \ref{LDF}. This formalism has been used to compute large deviation functions for other choices of $U(x)$ for Gaussian stationary Markov processes \cite{Majumdar2002b}. 
To proceed, we 
need to solve the following eigenvalue equation:
\begin{align}
\hat{H}_k\phi_{\lambda(k)}(x) = \lam(k) \phi_{\lam(k)}(x) \,,~~\label{ev-harm}
\end{align}
with specified boundary conditions, where the Hamiltonian is given by Eq.~\eqref{H-harm-loc}. Performing the transformations
\begin{align}
\sqrt{\frac{\kappa }{D}} \,x=u \,,~~\bar{\lam}(k)=\frac{\lam(k)}{\kappa }+\frac{1}{2} \,,~~\Phi_{\bar \lam}(u)=\frac{\kappa }{D} \phi_\lam(x) \,,~~\text{and}~~v=\sqrt{\frac{\kappa }{D}} \,y \,, \label{transformations}
\end{align}
the above equation becomes the well-known parabolic cylinder equation \cite{Magnus1966,wiki}:
\begin{align}
\partial_u^2 \Phi_{\bar \lam}(u)-\left( \frac{u^2}{4}+a\right)\Phi_{\bar \lam}(u) = \frac{k}{\sqrt{\kappa D}} \,\delta(u-v) \,\Phi_{\bar \lam}(u)~~\text{with}~~a=-\bar{\lambda} \,. \label{ev-diff-eq}
\end{align}
The two linearly independent solutions of the homogeneous part of the equation are given by 
\begin{align}
 &\mathscr{U}(a,u) = \mathscr{D}_{-a-1/2}(u),~~ \mathscr{V}(a, u ) =   \frac{\Gamma[a+1/2]}{\pi} \l( \sin(\pi a) \,\mathscr{D}_{-a-1/2}(u) +\mathscr{D}_{-a-1/2}(-u)  \r) . \label{gen-sol-PCE}
\end{align}
where $\mathscr{D}_{-a-1/2}(u)$ is the parabolic cylinder function.
In this case, the general solutions of Eq.~\eqref{ev-diff-eq} can also be written as 
\begin{align}
\Phi_{\bar \lam}(u)=
\begin{cases}
A_1\,\mathscr{D}_{\bar \lam -1/2}(u)+B_1\,\mathscr{D}_{\bar \lam -1/2}(-u) \,, & \text{for } u \ge v \\
A_2\,\mathscr{D}_{\bar \lam -1/2}(u)+B_2\,\mathscr{D}_{\bar \lam -1/2}(-u) \,, & \text{for } u <  v 
\end{cases}
\label{gen-sol}
\end{align}
The constants $A_{1,2}$ and $B_{1,2}$ are determined by imposing continuity at $u=v$,
\begin{align}
\Phi_{\bar \lam}(u)|_{u \to v^-}=\Phi_{\bar \lam}(u)|_{u \to v^+} \,, \label{continuity}
\end{align}
discontinuity of the first derivative at $u=v$,
\begin{align}
\left[ \partial_u \Phi_{\bar \lam}(u)\right]_{u \to v^+} - \left[ \partial_u \Phi_{\bar \lam}(u)\right]_{u \to v^-} 
= \frac{k}{\sqrt{\kappa  D}} \Phi_{\bar \lam}(v) \,, \label{D-discontinuity}
\end{align}
and the boundary conditions. In this section we consider two types of boundary conditions:
\begin{description}
\item[(A)] The particle is allowed to move in the whole space $-\infty < x < \infty $ under the influence of the harmonic potential.  The boundary conditions in this case are 
\begin{align}
\phi_\lam(x \to \pm \infty) \to 0~~\text{i.e.}~~\Phi_{\bar \lam}(u \to \pm \infty) \to 0 \,. \label{BC-full-space-harm}
\end{align}
\item[(B)] The particle is allowed to move in the semi-infinite space $0 \leq x < \infty$ under the influence of harmonic potential with a reflecting boundary at $x=0$. The boundary conditions in this case are 
\begin{align}
\partial_u\Phi_{\bar \lam}(u=0) = 0 \,,~~~\text{and}~~~\Phi_{\bar \lam}(u \to \infty) \to 0 \,. \label{BC-mixed-harm}
\end{align}
\end{description}
Let us first look at Case (A). 

\subsubsection{Case (A): On infinite line --}
\label{LDF-harm-IL}
\begin{figure}[t]
	\centering
	\includegraphics[width=\textwidth]{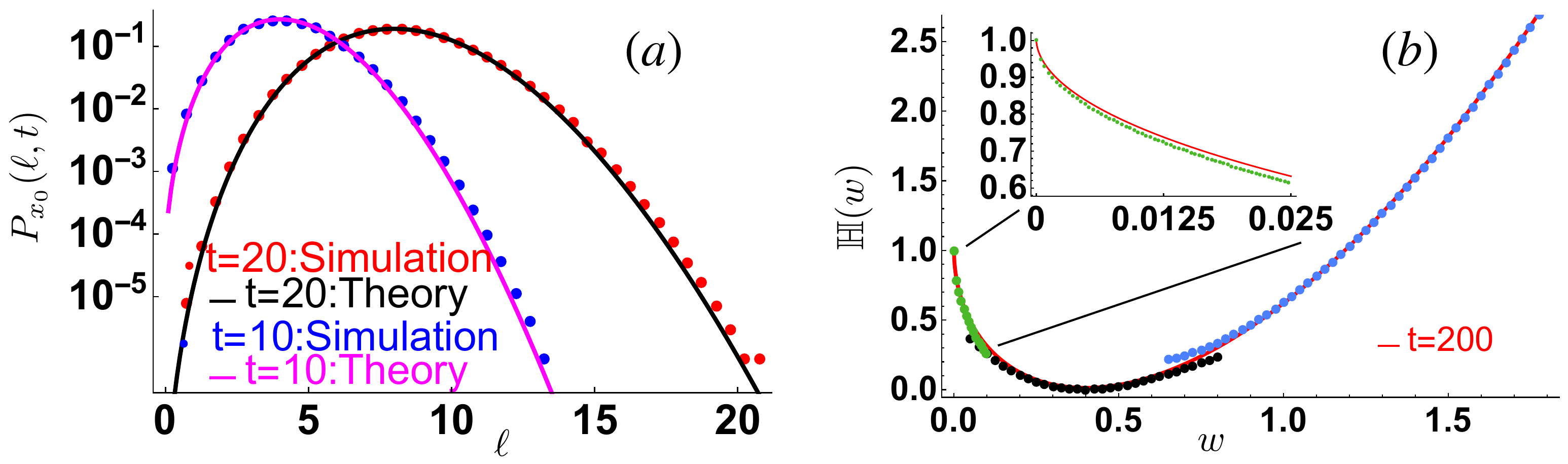}
	\caption{{\bf (a):} Distribution of the local time density spent, till times $t=10$ and $20$ at position $y=0$, by a Brownian particle moving in a harmonic potential over the full space. {\bf (b):} The asymptotic behaviour of the LDF $\mathbb{H}(w)$ in different regimes (given in Eq.~\eqref{LDF-asymp-harm-full}) is compared with the actual distribution given in Eq.~\eqref{P_U_t} for $t=200$. The behaviour at small $w$ is zoomed in the inset. For both the plots, $x_0=0$, $\kappa=1$ and $D=1$.  Numerical simulations were performed using a forward Euler Scheme with time-step $dt=10^{-3}$, and data were averaged over $10^7$ realisations.}
	\label{fig1p1}
\end{figure}

For this case, the boundary conditions in Eq.~\eqref{BC-full-space-harm} imply $B_1=0$ and $A_2=0$. Now, using the continuity at $u=v$, we get 
\begin{align}
\Phi_{\bar \lam}(u)=C
\begin{cases}
\mathscr{D}_{\bar \lam -1/2}(-v) \,\mathscr{D}_{\bar \lam -1/2}(u) \,, &\text{for } u\ge v \\
\mathscr{D}_{\bar \lam -1/2}(v) \,\mathscr{D}_{\bar \lam -1/2}(-u) \,, &\text{for } u <  v 
\end{cases}
\end{align}
where $C$ is a constant which can be fixed by requiring the eigenfunctions to be normalized. Inserting this expression for the eigenfunction in the discontinuity condition 
Eq.~\eqref{D-discontinuity}, we get
\begin{align}
-\left[ \frac{\mathscr{D}_{\lam/\kappa  +1}(v)}{\mathscr{D}_{\lam/\kappa }(v)}+ \frac{\mathscr{D}_{\lam/\kappa  +1}(-v)}{\mathscr{D}_{\lam/\kappa }(-v)}\right]=
\frac{\lam}{\kappa }~\left[ \frac{\mathscr{D}_{\lam/\kappa  -1}(v)}{\mathscr{D}_{\lam/\kappa }(v)}+ \frac{\mathscr{D}_{\lam/\kappa  -1}(-v)}{\mathscr{D}_{\lam/\kappa }(-v)}\right]=\frac{k}{\sqrt{\kappa D}} \,, \label{ev-equ-full}
\end{align}
solving which we can get $\lam$ for a given value of $k$. Note that this equation is exactly in the form of Eq.~\eqref{eg-sol-eq}. Solving for the smallest eigenvalue $\lam_0(k)$, and then performing the saddle point integration in Eq.~\eqref{ILT-mcalQ_p}, one finds the distribution Eq.~\eqref{P_U_t} for large $t$. Note that in this case the survival probability is $S(t|x_0)=1$ as the particle always survives. In Fig.~\ref{fig1p1}(a), we plot the distribution of the local time, for times $t=10$ and $20$, at position $y=0$ (i.e.\@ $v=0$) for a Brownian particle starting from $x_0=0$. 

From the structure of the saddle point integration in Eq.~\eqref{ILT-mcalQ_p}, it is clear that the distribution $P_{x_0}(\ell,t)$ has a large deviation form 
\begin{align}
P_{x_0}(\ell,t) \asymp \exp\!\left( -t~\mathbb{H}\left( \frac{\ell}{t}\right)\right), \label{LDF-locT-harm-full}
\end{align}
where  the large deviation function (LDF) $\mathbb{H}(w)$ is obtained through the following Legendre transformation 
\cite{Touchette2009}:
\begin{align}
&\mathbb{H}(w)= \max\limits_{k} \!\left( -k w+\lam_0(k)\right)= -k^*(w) \,w+\lam_0(k^*(w)) \,, \\ 
&\text{ where } \left(\frac{d \lam_0(k)}{dk}\right)_{k=k^*}=w=\frac{\ell}{t} \,. \label{LDF-H-free}
\end{align}
In mathematics literature, for diffusive processes, a general expression for the large deviation function $\mathbb{H}(w)$ of $w=\ell/t$ (known as the empirical distribution) has been obtained by Donsker and Varadhan in terms of the generator of the process \cite{Touchette2018, Angeletti2016,Touchette2009,Gartner1977}. 
From this result, one can in principle evaluate the  large deviation function of $\ell$ and also compute large deviation functions of other functionals like $T_t[x(\tau)]$ (in Eq.~\eqref{T_t[x]}) through the contraction principle \cite{Touchette2009}. However, here, we apply a different method, based on analyzing the behaviour of $\lam_0(k)$ in the $k \to 0^\pm$ and $k \to \pm \infty$ limits, to compute the large deviation function in different asymptotic limits {\it explicitly}.

Following this procedure, we find the following explicit asymptotic behaviours of the large deviation function
\begin{align}
\mathbb{H}(w) \approx {}&
\begin{dcases}
\kappa  - 2\kappa \left(2 w \sqrt{\frac{2D}{\kappa \pi}}\right)^{1/2} +  2 w \left(1 - \ln 2 \right) \sqrt{\frac{2D\kappa}{\pi}} +O(w^{3/2}) &\text{for }w \to 0 \\
\frac{(w-\mu)^2}{2\sigma^2} - \Omega \left(w-\mu\right)^3 + {O} \!\left(|w-\mu|^4\right)  &\text{for } \left|w-\mu\right| \leq \sigma \\
D \,w^2-\frac{\kappa}{2} +\frac{\kappa^{2}}{8 D w^{2}} +O(w^{-4}) &\text{for } w \to \infty
\end{dcases}
\label{LDF-asymp-harm-full}
\end{align}
where
\begin{equation}
	\sigma=\sqrt{\frac{\ln 2}{\pi  D}},\quad \mu = \sqrt{\frac{\kappa }{2 \pi D}},\quad 
	\Omega = 
	\frac{D^2\pi \sqrt{\pi} \left(36 \left(\ln 2\right)^2-\pi^2\right)}{48 \sqrt{2 D \kappa} \left(\ln 2\right)^3}.
	\label{mu-sigma-A3-definition}
\end{equation}
for $y=0$.
The asymptotic forms in Eq.~\eqref{LDF-asymp-harm-full} suggest that the typical fluctuations of $w=\ell/t$ are Gaussian with $\la w \ra = \la \ell_t/t\ra = \sqrt{{\kappa }/{2 \pi D}}$ and variance 
$\la (\Delta w)^2 \ra=\sigma^2=\ln 2/{\pi  D}$. This means one can write $w =\la w \ra  + \sigma~\chi$ for $|w- \la w \ra | \lesssim \sigma$ where $\chi$ is a mean zero, unit variance Gaussian variable. 

The asymptotic behaviours in Eq.~\eqref{LDF-asymp-harm-full} can be obtained as follows. First, we note that for $y=0$, Eq.~\eqref{ev-equ-full} simplifies to
\begin{align}
\frac{\mathscr{D}_{\lam/\kappa  +1}(v)}{\mathscr{D}_{\lam/\kappa }(v)} &= - \frac{1}{2}\frac{k}{\sqrt{\kappa D}}
\label{ev-equ-full-y_0-0}
\end{align}
where $\Gamma[x]$ is the gamma function.
Now, for $k=0$, it is easy to show that the solution for the smallest eigenvalue $\lam_0$ of the above equation  is zero.  Hence, in the $k \to 0^+$ limit, we expect $\lambda_0(k)$ should also go to zero. This suggests that we expand the left hand side of Eq.~\eqref{ev-equ-full-y_0-0} in powers $\nu=\lam_0/\kappa$ around $\nu=0$. We get
\begin{equation*}
\sqrt{2 \pi} \nu + \sqrt{2\pi} \ln 2~\nu^2 + \nu^{3} \sqrt{2\pi}\left(\frac{ (\ln 2)^{2}}{2} + \frac{\pi^2}{24}\right) + O(\nu^4)  =\frac{k}{\sqrt{\kappa D}} \,, \text{ with } \nu=\frac{\lam_0}{\kappa} \,. 
\end{equation*}
Solving the above order by order in small $k$, we get 
\begin{align}
\lam_0(k) \approx   \sqrt{\frac{\kappa }{2 \pi D}} k - \frac{\ln 2}{2 \pi  D}k^2 + k^{3} \left(  \frac{3 \sqrt{2} (\ln 2 )^{2}}{8 \pi^{\frac{3}{2}} D^{\frac{3}{2}} \sqrt{\kappa}} - \frac{\sqrt{2\pi} }{96 D^{\frac{3}{2}} \sqrt{\kappa}}\right) + O(k^4)  \text{ for } k \to 0^+ .
\end{align}
Using the above  expression in Eq.~\eqref{LDF-H-free}, we get 
\begin{align}
\mathbb{H}(w) &\approx  \frac{\left( w - \mu\right)^{2}}{2 \sigma^2} - \Omega \left( w - \mu \right)^{3} + {O} \!\left(|w-\mu|^4\right) 
\text{ for } \left|w- \mu \right| \leq \sigma \,, 
\end{align}
where $\mu$, $\sigma$, and $\Omega$ are defined in Eq.~\eqref{mu-sigma-A3-definition}.
Note that $k \to 0^+$ provides the LDF for $\ell \to \sqrt{\frac{\kappa }{2 \pi D}}~t$ from below. Now, let us look at the $k \to \infty$ limit along the real line. Note from Eq.~\eqref{ev-equ-full-y_0-0} that as $k \to \infty$, the left hand side also has to go to $\infty$ and that happens if $\lam/\kappa  \to 1$. Hence, we can look for a solution for $\lam_0$ of the form $\lam_0/\kappa  = 1 - \epsilon$, such that $\epsilon \to 0$ as $k \to \infty$. Inserting this form in Eq.~\eqref{ev-equ-full-y_0-0}, expanding in powers of $\epsilon$, and solving order by order for $\epsilon$, we get $\epsilon = \sqrt{\frac{2\kappa D}{\pi}}\frac{2}{k} - \frac{8\kappa D}{\pi k^2} \left(1-\ln 2\right) +O(k^{-3})$ which finally provides
\begin{align}
\lam_0(k) \approx \kappa(1-\epsilon)= \kappa  \left( 1 - 2\kappa \sqrt{\frac{2D}{\pi \kappa}} \, \frac{1}{k} + \frac{8\kappa D}{\pi k^2} \left(1-\ln 2\right) + O(k^{-3})\right) \text{ for } k \to \infty \,.
\end{align}
Once again, using this form of $\lam_0(k)$ in  Eq.~\eqref{LDF-H-free}, we get
\begin{align}
\mathbb{H}(w) \approx  \kappa  - 2\kappa \left(2 w \sqrt{\frac{2D}{\kappa \pi}}\right)^{1/2} + 2 w \left(1 - \ln 2 \right) \sqrt{\frac{2D\kappa}{\pi}} +O(w^{3/2})
\quad \text{ for } w \to 0 \,,
\end{align}
since $k \to \infty$ corresponds to the $w \to 0$ (i.e.\@ $\ell \to 0$ ) limit of the saddle point result in Eq.~\eqref{ILT-mcalQ_p}.

From the above analysis, we observe that the range $0<k<\infty$ of $\lam_0(k)$, provides information about the distribution $P_0(\ell,t)$ (or equivalently the LDF $\mathbb{H}(w)$) over the region $0< \ell < \sqrt{\frac{\kappa }{2 \pi D}}~t$. To obtain the LDF $\mathbb{H}(w)$ over the rest of the region  $ \sqrt{\frac{\kappa }{2 \pi D}}~t < \ell < \infty$, one needs to analytically continue the $\lam_0(k)$ to negative $k$. If we write $k =-|k|$ and $\lam= -|\lam|$, then we get 
\begin{align}
\sqrt{\frac{2D}{\kappa }} \, |\lam|\, \frac{ \Gamma\!\left[\frac{\kappa +|\lam|}{2\kappa }\right]}{\Gamma\!\left[\frac{2\kappa +|\lam|}{2\kappa }\right]}&=|k|\,. \label{ev-equ-full-y_0-0-ve}
\end{align}
Following the same procedure, we solve this equation in the $|k| \to 0$ and $|k| \to \infty$ limits and find 
\begin{align}
\lam(k) \approx 
\begin{dcases}
	\sqrt{\frac{\kappa }{2 \pi D}} k - \frac{\ln 2}{2 \pi  D}~k^2 + k^{3} \left(- \frac{\sqrt{2\pi} }{96 D^{\frac{3}{2}} \sqrt{\kappa}} + \frac{3 \sqrt{2} (\ln 2)^{2}}{8 \pi^{\frac{3}{2}} D^{\frac{3}{2}} \sqrt{\kappa}}\right) + O(k^4) &\text{for } |k| \to 0\\
	- \frac{k^2}{4 D} - \frac{\kappa}{2} + \frac{D \kappa^2}{2 k^{2}} + O(k^{-4}) &\text{for } |k| \to \infty\\
\end{dcases}
\end{align}
Now, substituting these asymptotic behaviours in Eq.~\eqref{LDF-H-free}, we get 
\begin{align}
\mathbb{H}(w) &\approx 
\begin{dcases}
\frac{\left( w - \mu\right)^{2}}{2 \sigma^2} - \Omega \left( w - \mu \right)^{3} + {O} \!\left(|w-\mu |^4\right)
& \text{for } \Big{|}w- \mu \Big{|} \leq \sigma \\
D \,w^2 - \frac{\kappa}{2} +\frac{\kappa^{2}}{8 D w^{2}} +O(w^{-4})  &\text{for } w \to \infty
\end{dcases}
\end{align}
where $\mu$, $\sigma$, and $\Omega$ are defined in Eq.~\eqref{mu-sigma-A3-definition}.
The asymptotic behaviours of $\mathbb{H}(w)$ in Eq.~\eqref{LDF-asymp-harm-full} are shown in Fig.~\ref{fig1p1}(b). 
The above procedure can be developed further to find the next order terms systematically in each regime in Eq.~\eqref{LDF-asymp-harm-full}. Similarly,  the above analysis, although it is presented for $y=0$, can be straightforwardly extended to other values of $y$.

\subsubsection{Case (B): Semi-infinite space with a reflecting boundary at $x=0$}
In this section, we consider the case in which the particle is moving on the positive real axis with a reflecting boundary at the origin. 
We study the distribution of the local time spent by the particle at the origin. 
The eigenfunctions $\phi_\lam$, in this case, satisfy the boundary conditions in Eq.~\eqref{BC-mixed-harm}. Following the same procedure as in case (A), we get 
\begin{align}
\Phi_{\bar \lam}(u)=C
\begin{cases}
	\mathscr{D}_{\bar \lam -1/2}(v)\left [\mathscr{D}_{\bar \lam -1/2}(u)+\mathscr{D}_{\bar \lam -1/2}(-u) \right ]&\text{for } 0<u <  v  \\
	\mathscr{D}_{\bar \lam -1/2}(u)\left [\mathscr{D}_{\bar \lam -1/2}(v)+\mathscr{D}_{\bar \lam -1/2}(-v) \right ] &\text{for } u\ge v \\
\end{cases}
\text{ with } \bar \lam=\frac{\lam}{\kappa }+\frac{1}{2} \,,
\end{align}
where  the constant $C$ is determined by requiring the eigenfunctions to be normalized.
The condition for finding the eigenvalues $\bar \lam$, in this case, turns out to be
\begin{align}
\frac{\mathscr{D}_{\bar \lam +1/2}(v)\mathscr{D}_{\bar \lam -1/2}(-v) )+\mathscr{D}_{\bar \lam +1/2}(-v)\mathscr{D}_{\bar \lam -1/2}(v) }{\mathscr{D}_{\bar \lam -1/2}(v)~[\mathscr{D}_{\bar \lam -1/2}(v)+\mathscr{D}_{\bar \lam -1/2}(-v)]} = - \frac{k}{\sqrt{\kappa D}}~\label{ev-eq-mixed}
\end{align}
where $\bar \lam =\lam/\kappa  +1/2$ and $v=\sqrt{\kappa /D}~y$ as given in Eq.~\eqref{transformations}. Once again, solving for the dominant eigenvalue $\lam_0(k)$, and then performing the saddle point integration in Eq.~\eqref{ILT-mcalQ_p} for large $t$, one finds the distribution given in Eq.~\eqref{P_U_t} which has the large deviation scaling form. 

In this case also, it is possible to find the behaviour of the large deviation function in various asymptotic limits.  
We focus on the local time at the boundary, i.e.\@ at $v=0$. Computation of the local time at the boundary, known as the boundary local time, is important for various surface-mediated phenomena \cite{Greb1,Greb2,Greb3}.
For $v=0$, equation \eqref{ev-eq-mixed} simplifies to
\begin{align}
\frac{\mathscr{D}_{\lam/\kappa  +1}(0)}{\mathscr{D}_{\lam/\kappa  }(0)} = - \frac{k}{\sqrt{\kappa D}} \,. \label{ev-eq-mixed-1}
\end{align}
Note that the eigenvalue equation for this case is same as that obtained in Eq.~\eqref{ev-equ-full-y_0-0} for the no-boundary case, but with $k$ being replaced by $2k$. 
This implies that the distributions of the local time $\ell(t)$ corresponding to the two boundary conditions (A) and (B) [see Eqs.~(\ref{BC-full-space-harm},~\ref{BC-mixed-harm})] are related by $P_0^{(B)}(\ell,t)=\frac{1}{2}P_0^{(A)}\left(\frac{\ell}{2},t\right)$. 
The reason behind such a relation is simple: by symmetry, if $x(\tau)$ is an OU process on the real line, then $|x(\tau)|$ is an OU process on the positive real axis with reflections at the origin, where the harmonic potential has its minimum. 
Hence, the amount of time spent at a point $y$ by the reflected process $|x(\tau)|$ is same as the total amount of time spent by the original process $x(\tau)$ at the points $y$~and~$-y$. 
Since we are computing the local time at the origin $y=0$ (where the reflecting boundary is placed), the local time spent at $y=0$ in case (A) is half the local time spent at $y=0$ in case (B). 
This can be understood by noting that for a given trajectory $\{x(\tau):0\leq \tau \leq t\}$, the amount of time spent within a box $[-\epsilon,\epsilon]$ of infinitesimal size $2 \epsilon$ in case (A) is same as the amount of time spent by the corresponding reflected trajectory $\{|x(\tau)|:0\leq \tau \leq t\}$ in the box $[0,\epsilon]$ in case (B). 
Hence, dividing this time by the length of the box in the respective cases, one sees that the local time (density) in case (B) is twice that in case (A).

\begin{figure}[t]
	\centering
	\includegraphics[width=\textwidth]{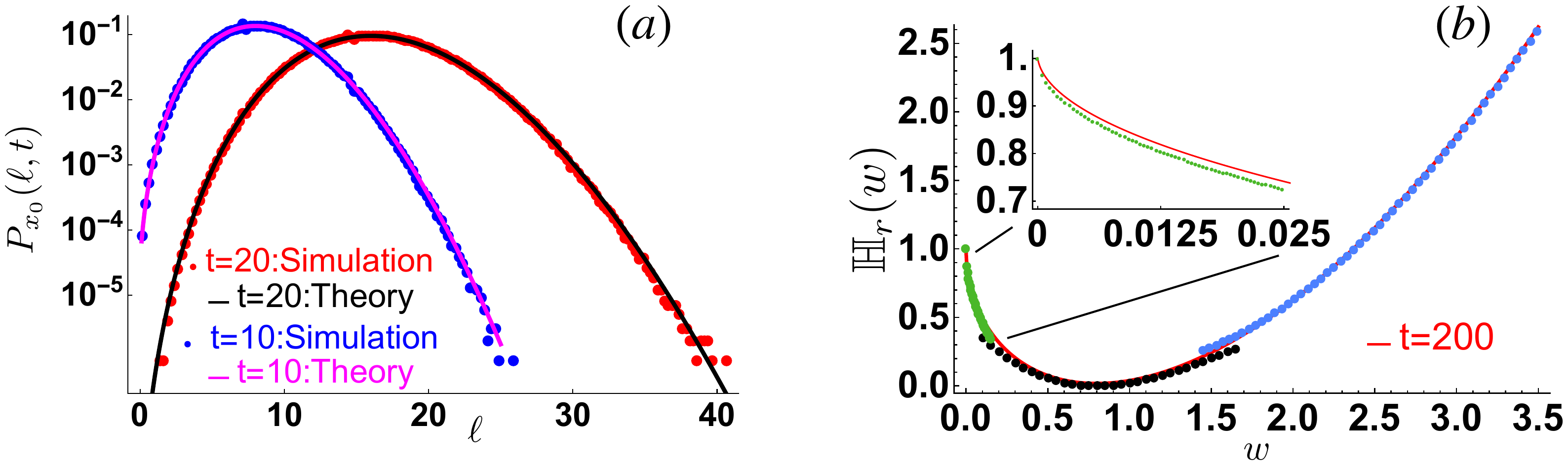}
	\caption{{\bf (a):} Distribution of the local time density spent, till times $t=10$ and $20$ at position $y=0$, by a Brownian particle moving in a harmonic potential on the positive real axis with a reflecting boundary at the origin. {\bf (b):} The asymptotic behaviour of the LDF $\mathbb{H}_{r}(w)$ of the distribution of the local time spent by the particle at $y=0$ till time $t=200$ in presence of a reflecting boundary at $x=0$. The solid red line is computed from Eq.~\eqref{P_U_t} and the points (solid discs) of different colours correspond to $\mathbb{H}(w)$ in different regimes as given in Eq.~\eqref{LDF-reflect}.  The parameters are $x_0=0, \kappa=1$ and $D=1$.  Numerical simulations were performed using a forward Euler Scheme with time-step $dt=5 \times 10^{-4}$, and data were averaged over $10^7$ realisations.}
	\label{fig2r}
\end{figure}

As a consequence, for the large deviation function $\mathbb{H}_{r}(w)$ with a purely reflecting boundary at the origin, one gets $\mathbb{H}_{r}(w)=\mathbb{H}\left(\frac{w}{2}\right)$. More explicitly, we get
\begin{align}
\mathbb{H}_{r}(w) \approx {}&
\begin{dcases}
\kappa  - 2\kappa \left(w\sqrt{\frac{2D}{\kappa \pi}}\right)^{1/2} + w\left(1 - \log 2\right) \sqrt{\frac{2D\kappa}{\pi}} + O(w^{3/2}) &\text{for } w \to 0 \\
\frac{(w-\mu_r)^2}{2\sigma_r^2} - \Omega_r (w-\mu_r)^3 + O\!\left(|w-\mu_r|^4\right) &\text{for } \left|w-\mu_r\right| \leq \sigma_r \\
\frac{D \,w^2}{4}-\frac{\kappa}{2} +\frac{\kappa^{2}}{2 D w^{2}}  + O(w^{-4}) &\text{for } w \to \infty
\end{dcases}
\label{LDF-reflect}
\end{align}
where,
\begin{equation}
\sigma_r=2\sigma=\sqrt{\frac{4\ln 2}{\pi  D}},\quad \mu_r = 2\mu=\sqrt{\frac{2\kappa }{\pi D}},\quad \Omega_r =\frac{\Omega}{8}= \frac{D^2\pi \sqrt{\pi} \left(36 \left(\ln 2\right)^2-\pi^2\right)}{384 \sqrt{2 D \kappa} \left(\ln 2\right)^3}
\end{equation}
These asymptotic behaviours of the LDF $\mathbb{H}_{r}(w)$ in the presence of a purely reflecting boundary at $x=0$, with $y=0$, are verified in Fig.~\ref{fig2r}.

It is important to mention that the relation $\mathbb{H}_{r}(w)=\mathbb{H}\left(\frac{w}{2}\right)$ holds only when the  the reflecting wall is placed at the origin, where the harmonic potential has its minimum; otherwise it breaks down. 
In such cases, one can follow the same procedure to solve the eigenvalue equation \eqref{ev-harm}, except that now the derivative boundary condition in Eq.~\eqref{BC-mixed-harm} has to be appropriately modified by equating the outward normal probability current across the boundary to zero in the corresponding Fokker-Planck equation.
Explicitly, if the reflecting boundary were placed at a position $x=r$, the relevant boundary conditions would be
\begin{equation}
	\dh_u \Phi_{\bar \lam} + \frac{u}{2} \Phi_{\bar \lam}  \Big|_{u = r\sqrt{\kappa/D}} = 0 \,,~~~\text{and}~~~\Phi_{\bar \lam}(u \to \infty) \to 0 \,.
\end{equation}

\subsection{Local time statistics not conditioned on survival with a fully absorbing boundary at the origin}
\label{LT-NC-S}
In the previous section, we studied the statistical properties of local time spent at location $y$ by the particle, conditioned on survival till time $t$ in presence of a fully absorbing boundary at the origin. As discussed 
in the introduction, often one needs to look at the statistics of $\ell_t(y)$ (or more generally of $T_t[x(\tau)]$) not conditioned on survival. In such cases, one needs to take into account the contribution to $\ell_t(y)$ from those trajectories which got absorbed before $t$, in addition to the contribution from those which are still surviving at time $t$. 
Let the MGF of $\ell_t(y)$ in this case be denoted by  $\mathbb{Q}(k,t \given x_0)$, where $x_0 >0$ is the initial position of the particle. Clearly, this quantity can be written as sum of two terms:
\begin{align}
\mathbb{Q}(k,t \given x_0) = Q(k,t \given x_0) + Q_{\text{fp}}(k,t \given x_0) \,, \label{mbb_Q}
\end{align}
where $Q(k,t \given x_0)$ represents the contribution from the trajectories surviving till time $t$ and is given by Eq.~\eqref{GF_1}. The second term $Q_{\text{fp}}(k,t \given x_0)$ represents the contribution from the trajectories which have already been absorbed at the origin before the observation time $t$. 
Of course, these trajectories contribute to $\ell_t(y)$ as long as they survive.
As a result, the moments can be written as the sum of two parts:
\begin{align}
\mean{\ell_t(y)^n} = \mean{\ell_t(y)^n}_{\rm s} + \mean{\ell_t(y)^n}_{\rm fp} \,, \label{ell_t(y)-sum}
\end{align}
where the first term with subscript `s' represents the contribution from the trajectories surviving till time $t$ (i.e.\@ the contribution from $Q(k,t|x_0)$) and the second term with subscript `fp' represents contribution from the trajectories absorbed before time $t$ (i.e.\@ the contribution from $Q_{\rm fp}(k,t|x_0)$).

In terms of the joint probability density $P_a(\ell,t_f|x_0)$ of the local time spent at $y$ till the first passage time $t_f$ (given that the particle started from $x_0$), the MGF $Q_{\text{fp}}(k,t|x_0)$ can be written as 
\begin{align}
Q_{\text{fp}}(k,t|x_0) =\int_0^\infty d\ell \e^{-k \ell}P_{\text{fp}}(\ell,t|x_0) \text{ where } P_{\text{fp}}(\ell,t|x_0)= \int_0^tdt_f\, P_a(\ell,t_f|x_0) \,. \label{Q_fp(k,t)-0}
\end{align}
Note that the subscript `fp' in $P_{\text{fp}}(\ell,t|x_0)$ denotes the distribution of local time $\ell_t(y)$ on the ensemble of trajectories which have crossed the origin before time $t$. In other words, $P_{\text{fp}}(\ell,t|x_0)d\ell$ represents the joint probability that the local time spent at location $y$ is within $\ell$ to $\ell +d\ell$ and $t_f \leq t$.

The Laplace transform of $P_a(\ell,t_f|x_0)$ with respect to both $\ell$ and $t_f$ is defined as
\begin{align}
\widetilde{P}_a(k,q|x_0)=  \int_0^\infty dt_f \e^{-q t_f}\int_0^\infty d\ell \e^{-k\ell} \, P_a(\ell,t_f|x_0) \,.
\label{tildeP_a}
\end{align}
It is easy to see from Eq.~\eqref{Q_fp(k,t)-0} that the Laplace transform of $Q_{\text{fp}}(k,t|x_0)$ with respect to time, denoted by 
$\widetilde{Q}_{\text{fp}}(k,q|x_0)$ is related to $\widetilde{P}_a(k,q|x_0)$ as
\begin{align}
\widetilde{Q}_{\text{fp}}(k,q|x_0) &= \int_0^\infty dt \e^{-q t}Q_{\text{fp}}(k,t|x_0) = \frac{\widetilde{P}_a(k,q|x_0)}{q} \,. \label{tilde_Q_fp}
\end{align}
Hence, at this stage, it is crucial to obtain $\widetilde{P}_a(k,q|x_0)$ explicitly. As shown in \ref{deri-tildeP_a}, this quantity satisfies the following differential equation:
\begin{equation}
	D\frac{\dh^2 }{\dh x_0^2}\LaplaceTransform{P}_a(k,q|x_0)
	- V'(x_0)\frac{\dh }{\dh x_0} \LaplaceTransform{P}_a(k,q|x_0)
	-(k \,\delta(x_0-y)+q) \LaplaceTransform{P}_a(k,q|x_0) = 0
	\label{diff-eq-tildeP_ab}
\end{equation}
with boundary conditions
\begin{align}
\begin{split}
	\lim_{x_0\to0} \LaplaceTransform{P}_a(k,q\given x_0) &= 1
	\\ \lim_{x_0\to\infty} \LaplaceTransform{P}_a(k,q\given x_0) &= 0
\end{split}
\label{tildeP_a-bc}
\end{align}
for $q >0$. For $x_0 \to 0$, the first passage time $t_f \to 0$ and $\ell \to 0$, which (from Eq.~\eqref{tildeP_a}) implies 
$\lim_{x_0\to0} \LaplaceTransform{P}_a(k,q\given x_0) = 1$. On the other hand, for $x_0 \to \infty$, the first passage time $t_f \to \infty$, which again (from Eq.~\eqref{tildeP_a}) implies $\lim_{x_0 \to \infty} \LaplaceTransform{P}_a(k,q\given x_0) = 0$.

\subsubsection[{Stationary distribution Ps(l|x0)}]{Stationary distribution $P_s(\ell | x_0)$ }
\label{sta-hp}
While the contribution to the cumulants from the first term $Q(k,t|x_0)$ in Eq.~\eqref{mbb_Q} decays exponentially with time,  the contribution from the second term saturates at large times. This implies the distribution of $\ell$ becomes stationary at $t \to \infty$.  Denoting this stationary distribution by $P_{\rm st}(\ell|x_0)$, we have
\begin{align}
P_{\rm st}(\ell|x_0) = P_{\text{fp}}(\ell,t \to \infty|x_0) = \int_0^\infty dt_f\, P_a(\ell,t_f|x_0) \,,
\end{align}
where the subscript `st' denotes stationary state.
The Laplace transform of $P_{\rm s}(\ell|x_0)$ with respect to $\ell$ is related to $\widetilde{P}_a(k,q|x_0)$ as 
\begin{align}
\widetilde{P}_{\rm st}(k|x_0) = \int_0^\infty d\ell \e^{-k\ell} P_s(\ell|x_0) = \widetilde{P}_a(k,0|x_0) \,. \label{tilde_p_ell_t_inf}
\end{align}
To compute $\widetilde{P}_s(k|x_0)$ we solve the following differential equation, obtained by putting $q=0$ in  Eq.~\eqref{diff-eq-tildeP_ab}:
\begin{equation}
	D\frac{\dh^2 }{\dh x_0^2}\LaplaceTransform{P}_{\rm st}(k|x_0)
	- V'(x_0)\frac{\dh }{\dh x_0} \LaplaceTransform{P}_{\rm st}(k|x_0)
	-k \,\delta(x_0-y) \LaplaceTransform{P}_{\rm st}(k|x_0) = 0 \,.
	\label{diff-eq-tildeP_s}
\end{equation}
The boundary conditions are 
\begin{align}
\begin{split}
\lim_{x_0\to0} \LaplaceTransform{P}_{\text{st}}(k\given x_0) &=1 \\ 
\lim_{x_0\to \infty} \LaplaceTransform{P}_{\text{st}}(k\given x_0) & < \infty~(\text{finite}) \,.
\end{split}
\label{BC-tildeP_st}
\end{align}
 The boundary condition at $x_0 =0$ is same as given in Eq.~\eqref{tildeP_a-bc} with $q=0$, i.e.\@ $\lim_{x_0\to0} \LaplaceTransform{P}_{\text{st}}(k\given x_0) = 1$ which can be understood as before.
 However the boundary condition $x_0 \to \infty$ is different. In this limit, although (as we have noted earlier) $t_f \to \infty$, the function $U(x)$, being $\delta(x-y)$ does not make the local time necessarily diverge or approach zero. Hence, the appropriate boundary condition is that $\LaplaceTransform{P}_{\text{st}}(k\given x_0)$ is finite for $x_0 \to \infty$.
Additionally, the presence of the delta function in Eq.~\eqref{diff-eq-tildeP_s} implies two extra conditions:
\begin{align}
\begin{split}
& \widetilde{P}_{\rm st}(k|x_0 \to y^+) = \widetilde{P}_{\rm st}(k|x_0 \to y^-)  \quad\text{ and} \\
& \frac{\dh \widetilde{P}_{\rm st}(k|x_0)}{\dh x_0}\bigg{\vert}_{x_0 \to y^+} - \frac{\dh \widetilde{P}_{\rm st}(k|x_0)}{\dh x_0}\bigg{\vert}_{x_0 \to y^-} = k \widetilde{P}_{\rm st}(k|y) \,. 
\end{split} 
\label{conti-dis-continuity_y}
\end{align}
The differential equation \eqref{diff-eq-tildeP_s} has been written and analysed in the general context of first passage Brownian functionals \cite{Majumdar2005,Majumdar2020}.
In the following, we solve the differential equation \eqref{diff-eq-tildeP_s} for two choices of $V(x)$: (i) free particle ($V(x)=0$) and (ii) OU particle ($V(x)=\kappa x^2/2$).

\customHead{Free particle ($V(x_0)=0$):} 
In this case, Eq.~\eqref{diff-eq-tildeP_s} becomes 
\begin{equation}
	D\frac{\dh^2 \LaplaceTransform{P}_{\rm st}(k|x_0)}{\dh x_0^2}
		= k \,\delta(x_0-y) \LaplaceTransform{P}_{\rm st}(k|x_0) \,.
\end{equation}
Solving the above with the boundary conditions in Eqs.~\eqref{BC-tildeP_st} and \eqref{conti-dis-continuity_y}, we get 
\begin{equation}
	\LaplaceTransform{P}_{\rm st}(k\given x_0) = 
	\begin{dcases}
		1-\frac{k x_0}{ky+D}  & x_0<y\\
		\frac{D}{ky+D} & x_0 \geq y \,.
	\end{dcases} \label{P_s-free}
\end{equation}
Taking the inverse Laplace transform (see Eq.~\eqref{tilde_p_ell_t_inf}), we get 
\begin{align}
P_{\rm st}(\ell\given x_0) = \delta(\ell) \l( 1 - \frac{\op{min}(x_0,y)}{y} \r) + \frac{D \op{min}(x_0, y)}{ y^2} \exp\l(-\frac{\ell D}{y}\r) .
\label{P_s(ell)-fp}
\end{align}
The stationary values of the mean and variance of $\ell$ at large time are given by 
\begin{align}
\begin{split}
\mean{\ell(y)}^{\rm (st)} &=  \frac{\op{min}(x_0, y)}{D} \,,  \\
\mean{\ell(y)^2}_c^{\rm (st)} &=  \frac{\op{min}(x_0, y)}{D^2} \l[ 2y - \op{min}(x_0, y) \r] .
\end{split}
\label{mean-var-fp}
\end{align}

\customHead{OU particle ($V(x_0)=\kappa x_0^2/2$):} 
In this case, Eq.~\eqref{diff-eq-tildeP_s} becomes 
\begin{equation}
	D\frac{\dh^2 \LaplaceTransform{P}_{\rm st}(k|x_0)}{\dh x_0^2}
	- \kappa x_0 \frac{\dh \LaplaceTransform{P}_{\rm st}(k|x_0)}{\dh x_0}
	= k \,\delta(x_0-y) \LaplaceTransform{P}_{\rm st}(k|x_0) \,.
\end{equation}
Solving the above with the boundary conditions in Eqs.~\eqref{BC-tildeP_st} and  \eqref{conti-dis-continuity_y}, we get 
\begin{align}
\LaplaceTransform{P}_{\rm st}(k \given x_0) =
\begin{dcases}
  1 - \frac{ k \op{\mathcal{K}}(y,x_0)}{1 + k \op{\mathcal{K}}(y,y)}  & x_0 < y \\
  \frac{1}{1 + k \op{\mathcal{K}}(y,y) } & x_0 \geq  y
\end{dcases}
\text{where } \op{\mathcal{K}}(y,z) = \sqrt{\frac{\pi}{2 \kappa D}} \e^{-\frac{\kappa y^2}{2 D}} \op{erfi}\!\l(z\sqrt{\frac{\kappa}{2D}}\r).
\label{P_s-hp}
\end{align}
Performing the inverse Laplace transform, we obtain
\begin{align}
P_{\rm st}(\ell|x_0) = \l(1 - \frac{\mathcal{K}(y,\op{min}(x_0,y))}{\mathcal{K}(y,y)}\r)\delta(\ell) + \frac{\mathcal{K}(y,\op{min}(x_0,y))}{\mathcal{K}^2(y,y)} \exp\!\l(-\frac{\ell}{\mathcal{K}(y,y)}\r) . 
\label{P_s(ell)-hp}
\end{align}
The stationary values of the mean and variance of $\ell$ at large time are then given by 
\begin{align}
		\mean{\ell(y)}^{\rm (st)} ={}& \op{\mathcal{K}}(y,\op{min}(x_0,y)) \,, \label{mean-hp}\\
		\mean{\ell(y)^2}_c^{\rm (st)} ={}&
		\op{\mathcal{K}}(y,\op{min}(x_0,y)) \l[ 2\op{\mathcal{K}}(y,y) - \mathcal{K}(y,\op{min}(x_0,y)) \r] .
\label{var-hp}
\end{align}
Note that for $\kappa \to 0$, the results in Eqs.~\eqref{P_s(ell)-hp}, \eqref{mean-hp} and \eqref{var-hp} match with the results of the free particle case (Eqs.~\eqref{P_s(ell)-fp} and \eqref{mean-var-fp}), as expected.

\subsubsection[{Computation of Qfp(k,q|x0) in Eq.~\eqref{tilde_Q_fp}}]{Computation of $\widetilde{Q}_{{\rm fp}}(k,q|x_0)$ in Eq.~\eqref{tilde_Q_fp}}
\label{sec:Q_fp}
In this section, we focus on  $Q_{\text{fp}}(k,t|x_0)$ which can be computed by solving the differential equation \eqref{diff-eq-tildeP_ab}.
We once again solve for two choices of $V(x)$: (i) free particle ($V(x)=0$) and (ii) Harmonic potential ($V(x)=\kappa x^2/2$).

\customHead{Free particle ($V(x_0)=0$):}  In this case, equation \eqref{diff-eq-tildeP_ab} becomes
\begin{equation}
	D\frac{\dh^2 }{\dh x_0^2}\LaplaceTransform{P}_a(k,q|x_0)
	-(k \,\delta(x_0-y)+q) \LaplaceTransform{P}_a(k,q|x_0) = 0 \,.
	\label{diff-eq-tildeP-ab-free}
\end{equation}
Solving the above with the boundary conditions in Eqs.~\eqref{tildeP_a-bc}, and then inserting that solution in Eq.~\eqref{tilde_Q_fp}, we get 
\begin{align}
\widetilde{Q}_{\text{fp}}(k,q|x_0) = \frac{1}{q}
	\begin{dcases}
	   \frac{\op{exp}\!\l({-\sqrt{\frac{q}{D}}x_0}\r) + \frac{k}{\sqrt{qD}} \op{exp}\!\l({-\sqrt{\frac{q}{D}}y}\r) \op{sinh}\!\l(\sqrt{\frac{q}{D}}(y-x_0)\r)}{1+\frac{k}{\sqrt{qD}} \op{exp}\!\l({-\sqrt{\frac{q}{D}}y}\r) \op{sinh}\!\l(\sqrt{\frac{q}{D}}y\r)}  & x_0<y \\
	   \frac{\op{exp}\!\l({-\sqrt{\frac{q}{D}}x_0}\r)}{1 + \frac{k}{\sqrt{qD}} \op{exp}\!\l({-\sqrt{\frac{q}{D}}y}\r) \op{sinh}\!\l(\sqrt{\frac{q}{D}}y\r)} & x_0 \geq y \,.
	\end{dcases}
	\label{tildeQ_fp_free}
\end{align}
Note that for $q \to 0$, we get $\widetilde{Q}_{\text{fp}}(k,q \to 0|x_0) = \frac{1}{q} \widetilde{P}_s(k|x_0)$ as expected from Eqs.~\eqref{tilde_Q_fp} and \eqref{tilde_p_ell_t_inf}, where $\widetilde{P}_s(k|x_0)$ is given in Eq.~\eqref{P_s-free}.

\customHead{OU particle ($V(x_0)=\kappa x_0^2/2$):}  In this case, equation \eqref{diff-eq-tildeP_ab} becomes
\begin{equation}
	D\frac{\dh^2 }{\dh x_0^2}\LaplaceTransform{P}_a(k,q|x_0)
		- \kappa x_0 \frac{\dh }{\dh x_0} \LaplaceTransform{P}_a(k,q|x_0)
	-(k \,\delta(x_0-y)+q) \LaplaceTransform{P}_a(k,q|x_0) = 0 \,.
	\label{diff-eq-tildeP_ab-hp}
\end{equation}
Defining the function $\widetilde{R}_a(k,q|x_0)$ as
\begin{align}
\LaplaceTransform{R}_a(k,q|x_0) = \e^{-\frac{\kappa x_0^2}{4 D}}\widetilde{P}_a(k,q|x_0) \,, \label{tildeP_a-tildeR_a-rela}
\end{align}
and performing the transformations $u_0= \sqrt{\frac{\kappa}{D}} x_0$ and $v_0 = \sqrt{\frac{\kappa}{D}} y$,  the above differential equation is converted to the parabolic cylinder equation, as in Eq.~\eqref{ev-diff-eq}:
\begin{align}
 \frac{\dh^2 }{\dh u_0^2}\LaplaceTransform{R}_a(k,q|u_0) -\l(\frac{u_0^2}{4} +a \r)\LaplaceTransform{R}_a(k,q|u_0) 
= \frac{k}{\sqrt{\kappa D}} \,\delta (u_0-v_0) \LaplaceTransform{R}_a(k,q|u_0)
\label{diff-eq-tildeR_ab-hp}
\end{align}
where $a= q/\kappa - 1/2$.
The general solution of this equation is
\begin{align}
\widetilde{R}_a(k,q|u_0) = 
\begin{cases}
A_1 \,\mathscr{U}(a,u_0) + B_1 \,\mathscr{V}(a,u_0) \,, & u_0 < v_0 \\
A_2 \,\mathscr{U}(a,u_0) + B_2 \,\mathscr{V}(a,u_0) \,, & u_0 \geq  v_0
\end{cases}
\end{align}
where $\mathscr{U}(a,u_0)$ and  $\mathscr{V}(a,u_0)$ are the two independent solutions of the parabolic cylinder equations given 
in Eq.~\eqref{gen-sol-PCE}. 
The integration constants $A_{1.2}$ and $B_{1,2}$ are determined from the boundary conditions in Eqs.~\eqref{tildeP_a-bc}, along with the continuity of $\widetilde{R}(k,q|u_0)$ and the discontinuity of its derivative 
(with respect to $u_0$) across $u_0=v_0$. These extra conditions appear due to the presence of the delta function in Eq.~\eqref{diff-eq-tildeR_ab-hp}. After getting $\widetilde{R}(k,q|u_0)$, we get $\widetilde{P}_a(k,q|x_0)$ from Eq.~\eqref{tildeP_a-tildeR_a-rela}. Inserting this solution  in Eq.~\eqref{tilde_Q_fp} we get
\begin{align}
&\widetilde{Q}_{\text{fp}}(k,q|x_0) = \nonumber \\
& \frac{\e^{\frac{u_0^2}{4}}}{q}
	\begin{cases}
 \frac{\l[\mathscr{U}(a,v_0)\mathscr{V}(a
+1,v_0)+\l(a+\frac{1}{2}\r)\mathscr{V}(a,v_0)\mathscr{U}(a
+1,v_0)\r] \mathscr{U}(a,u_0)~+~ \frac{k}{\sqrt{\kappa D}} \mathscr{U}(a,v_0)\l[\mathscr{V}(a,v_0)\mathscr{U}(a,u_0) - \mathscr{U}(a,v_0)\mathscr{V}(a,u_0)\r] }{\l[\mathscr{U}(a,v_0)\mathscr{V}(a
+1,v_0)+\l(a+\frac{1}{2}\r)\mathscr{V}(a,v_0)\mathscr{U}(a
+1,v_0)\r] \mathscr{U}(a,0)~+~ \frac{k}{\sqrt{\kappa D}} \mathscr{U}(a,v_0)\l[\mathscr{V}(a,v_0)\mathscr{U}(a,0) - \mathscr{U}(a,v_0)\mathscr{V}(a,0)\r] } \,,  \\
 \phantom{mmmmmmmmmmmmmmmmmmmmmmmmmmmmmmmmmmm}  u_0 < v_0
	    \\
	      \\
	   \frac{\l[\mathscr{U}(a,v_0)\mathscr{V}(a
+1,v_0)+\l(a+\frac{1}{2}\r)\mathscr{V}(a,v_0)\mathscr{U}(a
+1,v_0)\r] \mathscr{U}(a,u_0)}{\l[\mathscr{U}(a,v_0)\mathscr{V}(a
+1,v_0)+\l(a+\frac{1}{2}\r)\mathscr{V}(a,v_0)\mathscr{U}(a
+1,v_0)\r] \mathscr{U}(a,0)~+~ \frac{k}{\sqrt{\kappa D}} \mathscr{U}(a,v_0)\l[\mathscr{V}(a,v_0)\mathscr{U}(a,0) - \mathscr{U}(a,v_0)\mathscr{V}(a,0)\r] } \,,  \\ 
\phantom{mmmmmmmmmmmmmmmmmmmmmmmmmmmmmmmmmmm} u_0 \geq v_0 \,.
	\end{cases} 
\label{tildeQ_fp_harm}
\end{align}
Note that in the above, $a$ is a function of $q$, as defined after Eq.~\eqref{diff-eq-tildeR_ab-hp}.
In the $q \to 0$ limit, we use the properties given in \ref{relations} and get $\widetilde{Q}_{\text{fp}}(k,q \to 0|x_0) = \frac{1}{q} \widetilde{P}_s(k|x_0)$, as expected from Eqs.~\eqref{tilde_Q_fp} and \eqref{tilde_p_ell_t_inf} where $\widetilde{P}_s(k|x_0)$ is now given in Eq.~\eqref{P_s-hp}.

\subsubsection{Approach to the stationary values of the cumulants: time dependence at large $t$: }
The expressions in Eqs.~\eqref{P_s(ell)-hp}, \eqref{mean-hp}, and \eqref{var-hp}  provide the stationary values of the mean and variance of $\ell_t(y)$ for $V(x)=0$ and $V(x)=\kappa x^2/2$ respectively. 
To understand the approach to these stationary values, one needs to expand $Q_{\text{fp}}(k,t|x_0)$ and $Q(k,t|x_0)$, given in Eq.~\eqref{mbb_Q}, in powers of $k$. Using this expansion in Eq.~\eqref{moments_T}, one can compute the moments of $\ell_t(y)$. As mentioned earlier, the first term $Q(k,t|x_0)$ in Eq.~\eqref{mbb_Q} represents the contribution from trajectories surviving till time $t$.  This contribution to the MGF  has been computed in sec.~\ref{LT-harm} for the harmonic potential and in \ref{free-particle} for the free particle. The Laplace transform $\widetilde{Q}_{\rm fp}(k,t|x_0)$ of the second term $Q_{\rm fp}(k,t|x_0)$ has been obtained in the previous section \ref{sec:Q_fp} for both cases. In particular, to get the mean  of the local time, we find it convenient to first find the Laplace transform of the mean, defined as 
\begin{align}
\widetilde{\mean{\ell_t(y)}}(q) &= \int_0^\infty dt \e^{-q t} \mean{\ell_t(y)}, \label{tilde<l_t>}
\end{align}
from the expansion of the Laplace transform (with respect to $t$) of the MGF in powers of $k$. 
After, that we perform the inverse Laplace transform to get the mean in the time domain.
Below, we compute the mean local time $\ell_t(y)$ for the free particle and the harmonic case separately. 

\customHead{Free particle:} We expand $\tilde{Q}_{\rm fp}(k,q|x_0)$ in Eq.~\eqref{tildeQ_fp_free} and $\tilde{Q}(k,q|x_0)$ in Eq.~\eqref{eq:Qtilde_s_uc_free} in powers of $k$ to obtain the Laplace transforms of moments of $\ell_t(y)$.
In particular, for the mean we get
\begin{equation}
\resizebox{\textwidth}{!}{$
\widetilde{\mean{\ell_t(y)}}(q)  =
\begin{dcases}
\overbrace{\frac{1}{q\sqrt{q D}} \e^{-\sqrt{\frac{q}{D}}y} \l[\op{sinh}\!\l(\sqrt{\frac{q}{D}}y\r) \e^{-\sqrt{\frac{q}{D}}x_0}- \op{sinh}\!\l(\sqrt{\frac{q}{D}}(y-x_0)\r) \r]}^{\text{from}~\widetilde{Q}_{\rm fp}(k,q|x_0)} & \\ 
& \\
\underbrace{-  \frac{\l[2 \op{sinh}\!\l(\sqrt{\frac{q}{D}}(y-x_0)\r)\l(1- \e^{-\sqrt{\frac{q}{D}}y} \r) \r. 
+ \l.2 \op{sinh}\!\l(\sqrt{\frac{q}{D}}y\r) \l( \e^{-\sqrt{\frac{q}{D}}(y+x_0)}- \e^{-\sqrt{\frac{q}{D}}x_0}\r)\r]}{q\sqrt{4qD}} }_{\text{from}~\widetilde{Q}(k,q|x_0)} \,,& x_0 <y\\
& \\
\overbrace{\frac{1}{q\sqrt{q D}}  \e^{-\sqrt{\frac{q}{D}}y} \e^{-\sqrt{\frac{q}{D}}x_0}\op{sinh}\!\l(\sqrt{\frac{q}{D}}y\r)}^{\text{from}~\widetilde{Q}_{\rm fp}(k,q|x_0)}+  \overbrace{\frac{1}{q\sqrt{Dq}} \op{sinh}\!\l(\sqrt{\frac{q}{D}}y\r) \l(1- \e^{- \sqrt{\frac{q}{D}}y}\r)  \e^{- \sqrt{\frac{q}{D}}x_0}}^{\text{from}~\widetilde{Q}(k,q|x_0)} \,,& x_0 \geq y 
\end{dcases}%
$}
\end{equation}
One can, in principle, perform the inverse Laplace transform with respect to $q$ to get an explicit exact expression of $\mean{\ell_t(y}$. However, we are interested in the large $t$ limit, in which we find
\begin{align}
\mean{\ell_t(y)} \simeq \frac{\min(x_0,y)}{D}  - \frac{yx_0}{D\sqrt{\pi D t}} \,, \label{mean_ell_free-particle}
\end{align}
We note that the asymptotic value of the above is the same as in Eq.~\eqref{mean-var-fp}.
Following a similar procedure, one can compute higher order moments as well.

\customHead{OU particle:} The contribution of the surviving trajectories to the mean 
of $\ell_t(y)$ can be obtained from the expansion of $Q({\bf k},t|x_0)$ in Eq.~\eqref{psi_n^0(x)-IL-ab}, as 
\begin{align}
\mean{\ell_t(y)}_{\rm s} &=  t \!\!{\sum_{n=1,3,5\dots} \!\!\!\! \e^{-t\lambda_n^{(0)}}\lambda_n^{(1)}(y)g_n^{(0)}f_n^{(0)}(x_0)}
- \!\!\!{\sum_{n=1,3,5\dots}\!\!\!\! \e^{-t\lambda_n^{(0)}}[g_n^{(1)}(y)f_n^{(0)}(x_0)+g_n^{(0)}f_n^{(1)}(y|x_0)]} \,, 
\label{<l_t(y)> surv}
\end{align}
where the expressions for
$\lambda_n^{(0)}$, $\lambda_n^{(1)}$, $g_n^{(0)}$, $g_n^{(1)}$, $f_n^{(0)}(x_0)$, and $f_n^{(1)}(y|x_0)$ are given in Eqs.~(\ref{lam_n^1}--\ref{g_n^11}), 
which have to be computed using the eigenvalues and eigenfunctions given in Eq.~\eqref{psi_n^0(x)-IL-ab}. In the leading order for large $t$, we get 
\begin{align}
\mean{\ell_t(y)}_{\rm s} &\simeq  t \e^{-t\lambda_1^{(0)}}\lambda_1^{(1)}(y)g_1^{(0)}f_1^{(0)}(x_0)  \simeq t \e^{-\kappa t} \frac{2 \kappa^2}{\pi D^2}  x_0 y^2 \e^{-\frac{\kappa y^2}{2D}} \,.
\end{align}
To find the contribution $\mean{\ell_t(y)}_{\rm fp}$ from the trajectories absorbed before time $t$, we expand $\widetilde{Q}_{\text{fp}}(k,q|x_0)$ in Eq.~\eqref{tildeQ_fp_harm} to linear order in $k$, and then take the first derivative with respect to $k$ at $k \to 0$. Adding this contribution to $\mean{\ell_t(y)}_{\rm s}$, we find that at large $t$ the leading time dependent contribution $t  \e^{-\kappa t}$ in $\mean{\ell_t(y)}_{\rm s}$ gets exactly cancelled and one finally gets 
\begin{align}
\mean{\ell_t(y)} \simeq \mean{\ell_t(y)}^{{\rm (st)}} - \mathbb{C}(x_0,y,\kappa,D) \e^{-\kappa t} 
\label{mean-ell_HP}
\end{align}
for both $x_0>y$ and $x_0 \leq y$, where $\mathbb{C}$ and $\mean{\ell_t(y)}^{{\rm (st)}}$ are time-independent constants.

The cancellation mentioned above actually occurs for all the lower order terms $t \e^{-n \kappa t}$ with $n=3,5,7,\dots$. This can be seen properly through an alternative formulation of the total MGF $\mathbb{Q}(k,t \given x_0)$ (defined in Eq.~\eqref{mbb_Q}). 
Next, we apply this method to a general confining potential $V(x)$ and obtain an expression for $\mathbb{Q}(k,t \given x_0)$. 
We then show that our results for the free and harmonic potentials (Eqs.\@ \eqref{mean_ell_free-particle} and \eqref{mean-ell_HP} respectively) are consistent with those obtained from this general expression.

\subsubsection{General method:}
\label{general-method}
In order to compute $Q_{\rm fp}(k,t \given x_0)$, we define the MGF $Q_{t'}(k,t|x_0)$ at time $t$ which involves contribution of the trajectories which have survived till time $t' <t$, i.e.\@ it does not include contributions from trajectories that have crossed the origin before time $t'$. Note that by definition, $Q_t(k,t|x_0) = Q(k,t|x_0)$. Hence, we have 
\begin{align}
	Q_t(k,t \given x_0)=Q(k,t|x_0) = \int_0^\infty \d x \bra{x} \e^{-\hat{H}_kt}\ket{x_0} \e^{-\frac{V(x)}{2D}+\frac{V(x_0)}{2D}}
	\label{Q_t(k,t)}
\end{align}
where ${\hat{H}_k}$ is given in Eq.~\eqref{H_p} with eigenfunctions satisfying the boundary conditions $\psi_n(x) \to 0$ as $x \to  \infty$ and $\psi_n(x=0)=0$.
The absorbing boundary condition $\psi_n(x)=0$ at $x=0$ can equivalently be expressed by assuming that the external potential $\mathcal{U}(x)$ is described by
\begin{equation}
\mathcal{U}_a(x) = 
		\begin{dcases}
			\infty &,\, x<0\\
			 \frac{V'(x)^2}{4 D}-\frac{V''(x)}{2} &,\, x\ge0 \,.
		\end{dcases}
\end{equation}
Still working with the ensemble of paths not absorbed till time $t$, we consider the MGF after a small interval $\Delta t$. The relevant MGF, $Q_t(k,t+\Delta t \given x_0)$, is given by
\begin{align}
	Q_t(k,t+\Delta t \given x_0) = \int_{-\infty}^\infty \d z \int_0^\infty \d x \bra{z} \e^{-\hat{\mathcal{H}}_k\Delta t}\ket{x}\bra{x} \e^{-\hat{H}_kt}\ket{x_0} \e^{-\frac{V(z)}{2D}+\frac{V(x_0)}{2D}}
\end{align}
where $\hat{\mathcal{H}}_k$ is a Hamiltonian similar to $\hat{H}_k$, but without the infinite potential over the negative real line, i.e.\@ it is a Hamiltonian over full space with eigenfunctions satisfying the boundary conditions $\psi_n(x) \to 0$ as $x \to  \pm \infty$. In the above, we have implicitly assumed that the time interval $\Delta t$ is small enough that we don't have to worry about multiple crossings of the origin within that interval. Note that $Q_t(k,t+\Delta t \given x_0)$ represents the MGF at time $(t+\Delta t)$, which includes contributions only from trajectories that have survived till time $t$; we do not worry about whether they get absorbed between $t$ and $(t+\Delta t)$.

On the other hand, we have
\begin{align}
	Q_{t+\Delta t}(k,t+\Delta t \given x_0) &= \int_0^\infty \d z \bra{z} \e^{-\hat{H}_k(t+\Delta t)}\ket{x_0} \e^{-\frac{V(z)}{2D}+\frac{V(x_0)}{2D}} \\
	&= \int_{0}^\infty \d z \int_0^\infty \d x \bra{z} \e^{-\hat{H}_k\Delta t}\ket{x}\bra{x} \e^{-\hat{H}_kt}\ket{x_0} \e^{-\frac{V(z)}{2D}+\frac{V(x_0)}{2D}} \,,
\end{align}
which represents the MGF at time $(t+\Delta t)$ that only includes contributions from trajectories surviving till time $(t+\Delta t)$. 
As a result, the difference
\begin{align}
	\begin{split}
		Q_t(k,t+\Delta t \given x_0) -{}& Q_{t+\Delta t}(k,t+\Delta t \given x_0) 
		\\={}& \int_{-\infty}^\infty \d z \int_0^\infty \d x \bra{z} \e^{-{\hat{\mathcal{H}}_k}\Delta t}\ket{x}\bra{x} \e^{-\hat{H}_kt}\ket{x_0} \e^{-\frac{V(z)}{2D}+\frac{V(x_0)}{2D}}
		\\{}& - \int_{0}^\infty \d z \int_0^\infty \d x \bra{z} \e^{-\hat{H}_k\Delta t}\ket{x}\bra{x} \e^{-\hat{H}_kt}\ket{x_0}\e^{-\frac{V(z)}{2D}+\frac{V(x_0)}{2D}}
	\end{split}
\end{align}
represents the contribution to the MGF from those trajectories which have been absorbed at the origin (i.e.\@ at the absorbing boundary) in the time interval $t$ to $t+\Delta t$. Taking the $\Delta t \to 0$ limit on both sides, we get 
\begin{align}
	\begin{split}
		\frac{d Q_{\rm fp}(k,t|x_0)}{dt} = & \int_{0}^\infty \d z \bra{z}{\hat{H}_k} \e^{-\hat{H}_kt}\ket{x_0} \e^{-\frac{V(z)}{2D}+\frac{V(x_0)}{2D}}
		\\ & - \int_{-\infty}^\infty \d z \int_0^\infty \d x \bra{z}{\hat{\mathcal{H}}_k}\ket{x}\bra{x} \e^{-\hat{H}_kt}\ket{x_0} \e^{-\frac{V(z)}{2D}+\frac{V(x_0)}{2D}}
	\end{split}
\end{align}
which provides $Q_{\rm fp}(k,t|x_0)$ upon integration over time from $0$ to $t$. We get 
\begin{align}
\begin{split}
Q_{\rm fp}(k,t|x_0) ={}& 1 - \overbrace{ \int_{0}^\infty \d z \bra{z}\e^{-\hat{H}_kt}\ket{x_0} \e^{-\frac{V(z)}{2D}+\frac{V(x_0)}{2D}} }^{Q(k,t|x_0)} 
\\
&- \int_0^t d\tau \int_{-\infty}^\infty \d z \int_0^\infty \d x \bra{z}{\hat{\mathcal{H}}_k}\ket{x}\bra{x} \e^{-\hat{H}_k\tau}\ket{x_0} \e^{-\frac{V(z)}{2D}+\frac{V(x_0)}{2D}} ,
\end{split}
\end{align}
 which implies 
 \begin{align}
 \mathbb{Q}(k,t|x_0) =1- \int_0^t d\tau \int_{-\infty}^\infty \d z \int_0^\infty \d x \bra{z}{\hat{\mathcal{H}}_k}\ket{x}\bra{x}\e^{-\hat{H}_k \tau}\ket{x_0} \e^{-\frac{V(z)}{2D}+\frac{V(x_0)}{2D}} .
 \label{mbbQ_final-form}
 \end{align}
Note that $ \mathbb{Q}(k,t \to 0|x_0) =1$ as expected. 
Next, we compute $ \mathbb{Q}(k,t |x_0)$ for two cases, a free particle and an OU particle, as before.

\begin{figure}[t]
	\centering
	\includegraphics[width=0.5\textwidth,keepaspectratio=true]{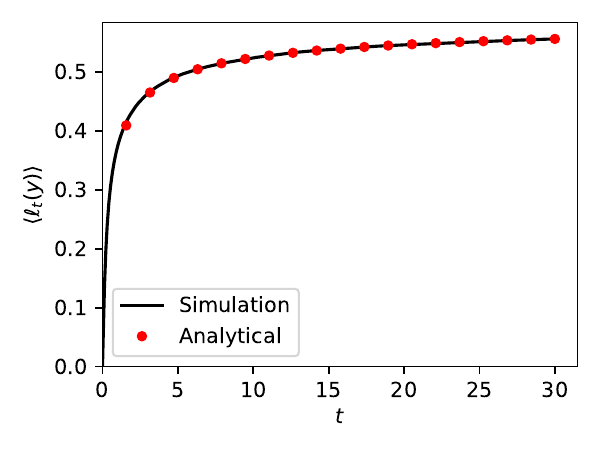}%
	\includegraphics[width=0.5\textwidth,keepaspectratio=true]{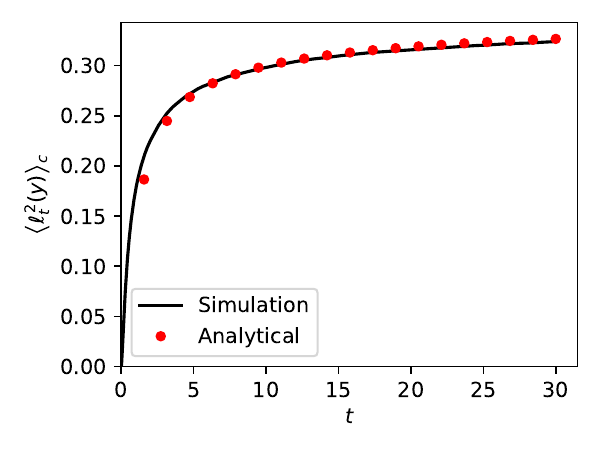}
	\caption{Numerical verification of the expressions of the mean ($\langle \ell_t(y) \rangle$) and variance ($\langle \ell_t^2(y) \rangle_c$) of the local time given in 
	Eqs.~\eqref{mean-lt-fp} and \eqref{loct-free-BM-variance} unconditioned on survival of a free particle with an absorbing boundary at the origin. In both cases, we have taken $x_0 = 0.5$, $y=0.3$, and $D=1/2$. For the simulations, we used a forward Euler scheme with time-step $dt= 10^{-7}$, and took $10^5$ realizations. }
	\label{<l_t(y)>-free-calc}
\end{figure}

 \customHead{Free particle:} In this case, $V(x)=0$, and consequently  $\bra{z}\hat{\mathcal{H}}_k\ket{z'} = -D\frac{d^2}{d z^2} \delta(z-z') + k\delta(y-z)\delta(z-z')$. Substituting this in Eq.~\eqref{mbbQ_final-form}, we get 
 \begin{align}
 \begin{split}
  \mathbb{Q}(k,t|x_0) &=1- k\int_0^t d\tau \bra{y} \e^{-\hat{H}_k \tau}\ket{x_0} \\
  &= 1- k \int_0^t d\tau \bra{y} \e^{-\hat{H}_0 \tau}\ket{x_0} + O(k^2)
  \end{split}
  \label{mbb_Q-free-particle}
 \end{align}
 where the free particle propagator $\bra{y}e^{-\hat{H}_0 \tau}\ket{x_0}$ in the presence of an absorbing boundary at $x=0$ is given by 
 \begin{align}
\bra{y}e^{-\hat{H}_0 \tau}\ket{x_0} = \frac{1}{\sqrt{4 \pi D \tau}}\l ( \e^{-\frac{(y-x_0)^2}{4 D \tau}} - \e^{-\frac{(y+x_0)^2}{4 D \tau}}\r) . \label{propagator}
 \end{align}
 From the coefficient of the term at $O(k)$ in Eq.~\eqref{mbb_Q-free-particle}, we correctly reproduce the result in Eq.~\eqref{mean_ell_free-particle}, i.e.\@
 \begin{align}
 \begin{split}
 \mean{\ell_t(y)} &= \int_0^t d\tau \bra{y} \e^{-\hat{H}_0 \tau}\ket{x_0} \\
 &\simeq \frac{\min(x_0,y)}{D} - \frac{yx_0}{D\sqrt{\pi D t}} \,\text{ for large $t$.}
 \end{split}
 \label{mean-lt-fp}
 \end{align}
It turns out that in this case one can compute the Laplace transform of all higher order moments $\widetilde{\mean{\ell_t^n(y}}(s)$ explicitly from which one can find their time dependence by performing inverse Laplace transforms. This calculation is presented in Appendix \ref{free-particle-hom}. In particular, we get the following explicit expression for the variance for large $t$:
\begin{align}
\begin{split}	
\left<l_t^2(y)\right>_c &=  \left<l_t^2(y)\right> - \left<l_t(y)\right>^2, ~~\text{where}\\
\left<l_t^2(y)\right>_{t\gg 1} &\approx \frac{2y\op{min}(y,x_0)}{ D^2} - \frac{2 y^2 \left( x_0 + \op{min}(y,x_0) \right)}{D^2\sqrt{\pi D t}} + O(t^{-1}), 
\end{split}
\label{loct-free-BM-variance}
\end{align}
In figure \ref{<l_t(y)>-free-calc}, we show that the obtained expressions for the mean and variance match well with simulations.

 \begin{figure}[t]
	\centering
	\includegraphics[width=0.5\textwidth,keepaspectratio=true]{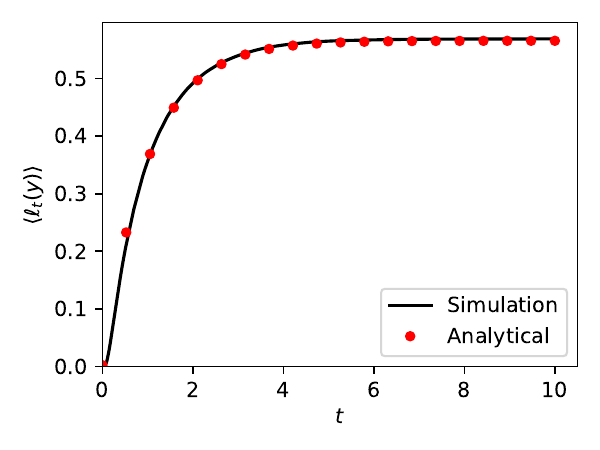}%
	\includegraphics[width=0.5\textwidth,keepaspectratio=true]{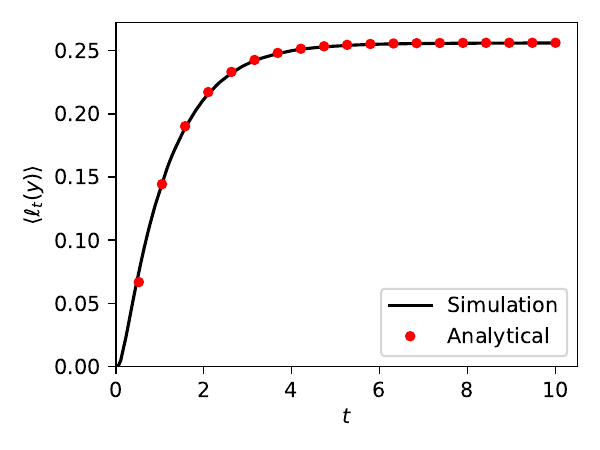}
	\caption{Numerical verification of the analytical expression of the mean local time $\langle \ell_t(y) \rangle$ of an OU particle unconditioned on survival, given in Eq. \eref{eq:<l_t(y)>-harm-alt-calc}, for $y=0.3$ (left) and $y=1.5$ (right). In both cases, we have taken $x_0=0.9$, $\kappa=1$, and $D=1/2$. For the simulations, we used a forward Euler scheme with time-step $dt=10^{-5}$, and took $10^5$ realizations.}
	\label{<l_t(y)>-harm-alt-calc}
\end{figure}
 
\customHead{OU particle: } To avoid possible confusion, in this case, let us denote the eigenvalues and eigenfunctions of ${\hat{H}_k}$ as $\lambda_n(k)$, $\psi_{n,k}(x)$ respectively, and of ${\hat{\mathcal{H}}_k}$ as $\mu_n(k)$, $\phi_{n,k}(x)$ respectively. Note that $\mu_n(0)$ and $\phi_{n,0}(x)$ are given in Eqs.~\eqref{psi_n^0(x)-IL} and  $\lambda_n(0)$ and $\psi_{n,0}(x)$ are given in Eqs.~\eqref{psi_n^0(x)-IL-ab}. 
Using the relevant completeness relations for $\psi_{n,k}(x)$ and $\phi_{n,k}(x)$ and performing the integration over time in Eq.~\eqref{mbbQ_final-form}, we get
\begin{multline}
		\mathbb{Q}(k,t|x_0) =  1
- \sum_{\mathclap{{\substack{n=1,3,5\dots\\m=0,1,2\dots}}}} \frac{\mu_m(k)[1- \e^{-\lambda_n(k) t}]}{\lambda_n(k)}  
 \\ 
\times \int_{-\infty}^\infty \d z \int_0^\infty \d x\,
\phi_{m,k}(z)\phi_{m,k}^*(x)\psi_{n,k}(x)\psi_{n,k}^*(x_0) \e^{-\frac{\kappa z^2}{4D}+\frac{\kappa x_0^2}{4D}}
	\label{mbbQ(k,t)}
\end{multline}
It is straightforward  to check that $\mathbb{Q}(0,t|x_0) = 1$ for $x_0>0$. We observe that the MGF has the  following form:
\begin{align}
\mathbb{Q}(k,t|x_0) ={}& \mathbb{Q}_\infty(k|x_0) + \mathbb{Q}_d(k,t|x_0)\label{Q_inf+Q_d}
\end{align}
where
\begin{align}
\begin{split}
\mathbb{Q}_\infty(k|x_0) ={}&  1
- \!\!\!\sum_{\substack{n=1,3,5\dots\\m=0,1,2\dots}}\!\! \frac{\mu_m(k)}{\lambda_n(k)} \int_{-\infty}^\infty \!\!\d z \int_0^\infty \!\!\d x\, \phi_{m,k}(z)\phi_{m,k}^*(x)\psi_{n,k}(x)\psi_{n,k}^*(x_0) \e^{-\frac{\kappa z^2}{4D}+\frac{\kappa x_0^2}{4D}} , \\
\mathbb{Q}_d(k,t|x_0)={}& \!\!\!\sum_{\substack{n=1,3,5\dots\\m=0,1,2\dots}}\!\!\!\! \frac{\mu_m(k) \e^{-\lambda_n(k) t}}{\lambda_n(k)} \int_{-\infty}^\infty \!\!\d z \int_0^\infty \!\!\d x\, \phi_{m,k}(z)\phi_{m,k}^*(x)\psi_{n,k}(x)\psi_{n,k}^*(x_0) \e^{-\frac{\kappa z^2}{4D}+\frac{\kappa x_0^2}{4D}} . \label{Q_d(k,t)}
\end{split}
\end{align}
Note that the the first term with subscript $\infty$ in Eq.~\eqref{Q_inf+Q_d} is actually $\mathbb{Q}_\infty(k|x_0)=\mathbb{Q}(k, t \to \infty|x_0) $. The second term (with subscript $d$) consists of terms decaying exponentially in time. 

\begin{figure}[t]
	\centering
	\includegraphics[width=0.5\textwidth,keepaspectratio=true]{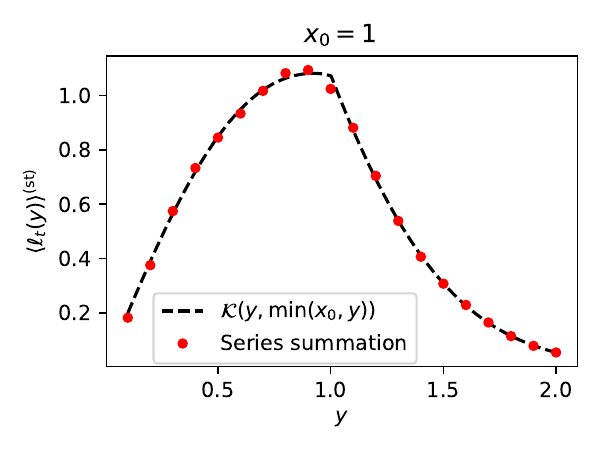}%
	\includegraphics[width=0.5\textwidth,keepaspectratio=true]{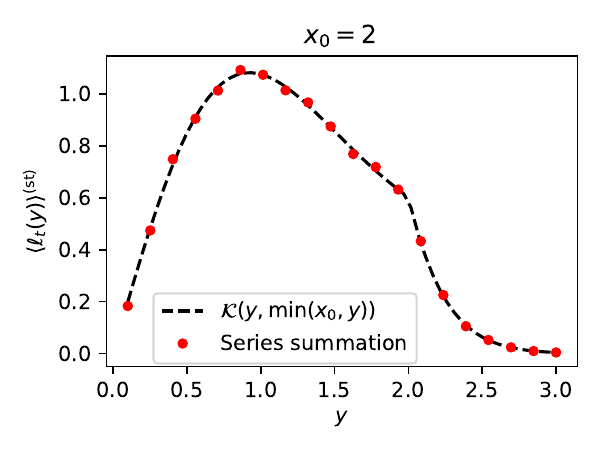}
	\caption{Numerical verification of the fact that the analytical expression of  $\langle \ell_t(y) \rangle^{\text{(st)}}$ in Eq.~\eqref{<l_t(y)>_st-series} converges to the closed form expression $\mathcal{K}(y,\op{min}(x_0,y))$ in Eq.~\eqref{P_s-hp} for two values of $x_0$. We set $\kappa=1$ and $D=1/2$.}
	\label{<l_t(y)>_st-validation}
\end{figure}

This structure suggests that all the cumulants of $\ell_t(y)$ become time-independent at large times.
This is expected because the survival probability decreases with increasing time.
As a result, the contribution to $\ell_t(y)$ starts saturating with time. To compute the cumulants, one can expand  $\mathbb{Q}_\infty(k,t|x_0)$ and $\mathbb{Q}_d(k,t|x_0)$ in Eq.~\eqref{mbbQ(k,t)} in powers of $k$ using the perturbation expansions of the eigenvalues and eigenfunctions as described in sec.~\ref{MGF}. The expansion of $\mathbb{Q}_\infty(k,t|x_0)$ would provide stationary values of the  cumulants of $\ell_t$. Recall that in sec.~\ref{sta-hp} we have computed stationary values of the mean and the variance using a different approach. 

We now focus on computing the time dependence of the mean at  large time. To get that, we expand $\mathbb{Q}_d(k,t|x_0)$ in Eq.~\eqref{Q_d(k,t)} to linear order in $k$ for large time $t$. We get 
\begin{align}
&\mathbb{Q}_d(k,t|x_0) \simeq k~\l(\frac{2 \pi D}{\kappa} \r)^{1/4} \e^{\frac{\kappa x_0^2}{4 D}}\sum_{n=1,3,5\dots} \e^{-n\kappa t} \frac{\psi_n^{(0)^*}(x_0) }{n\kappa}
\bigg{[}\mu_0^{(1)} \mathcal{J}_{0,n} \nonumber \\ 
& \qquad \qquad \qquad \qquad \qquad \qquad \qquad \qquad
 + \sum_{m=1,2,3\dots}\mathcal{J}_{m,n} \phi_m^{(0)}(y)\phi_0^{(0)^*}(y) \bigg{]} +O(k^2) \,, \nonumber \\ 
\begin{split}
&\simeq k~\l(\frac{2 \pi D}{\kappa} \r)^{1/4} \e^{\frac{\kappa x_0^2}{4 D}}\sum_{n=1,3,5\dots} \e^{-n\kappa t} \frac{\psi_n^{(0)^*}(x_0) }{n\kappa}~
\l[\sum_{m=0,1,2\dots} \mathcal{J}_{m,n} \phi_m^{(0)}(y)\phi_0^{(0)^*}(y) \r] 
\\ 
&  \qquad \qquad \qquad \qquad \qquad  \qquad \qquad \qquad \qquad \qquad \qquad \qquad \qquad \qquad ~
+O(k^2) \,,
\end{split}
\label{Q_d_lin-k}
\\
&\text{where, } \mathcal{J}_{m,n} = \int_0^\infty \psi_n^{(0)}(x) \phi_m^{(0)^*}(x) dx = \frac{1}{\sqrt{2^m m!}} \frac{\sqrt{2}}{\sqrt{2^n n! \pi}}\int_0^\infty \e^{-u^2} H_m(u)H_n(u) \,du \,. \label{mcalJ}
\end{align}
From this, one can readily find the approach of $\ell_t(y)$ to the stationary part given in Eq.~\eqref{mean-hp}. Adding the decaying part to the stationary part and then taking derivative with respect to $k$ at $k=0$, we get
\begin{align}
\mean{\ell_t(y)} &=
\l(\frac{2 \pi D}{\kappa} \r)^{1/4} \e^{\frac{\kappa x_0^2}{4 D}}\sum_{\substack{n=1,3,5\dots\\m=0,1,2\dots}}\l(1- \e^{-n\kappa t} \r) \frac{\psi_n^{(0)^*}(x_0) }{n\kappa}~
\l[\mathcal{J}_{m,n} \phi_m^{(0)}(y)\phi_0^{(0)^*}(y) \r] ,
\label{mean-lt-hp}
\end{align}
For large $t$, the above simplifies to
\begin{align}
\begin{split}
\ell_t(y) \approx\mean{\ell_t(y)}^{\rm (st)} - \l(\frac{2 \pi D}{\kappa} \r)^{1/4}\frac{\e^{-\kappa t}}{\kappa} \e^{\frac{\kappa x_0^2}{4 D}}~\psi_1^{(0)^*}(x_0) 
&\l[\sum_{m=0,1,2\dots} \mathcal{J}_{m,1} \phi_m^{(0)}(y)\phi_0^{(0)^*}(y) \r] \\
&~~~~~~~~~~~~~~~~~~~~~~~~ + O(\e^{-3 \kappa t })
\end{split}
\label{eq:<l_t(y)>-harm-alt-calc}
\end{align}
which is in the form given in Eq.~\eqref{mean-ell_HP}.
Note that in this method, we get an alternative expression for $\mean{\ell_t(y)}^{\rm (st)}$, apparently different from that given in Eq.~\eqref{mean-hp}:
\begin{align}
\mean{\ell_t(y)}^{\rm (st)} = \l(\frac{2 \pi D}{\kappa} \r)^{1/4} \e^{\frac{\kappa x_0^2}{4 D}}\sum_{n=1,3,5\dots} \frac{\psi_n^{(0)^*}(x_0) }{n\kappa}~
\l[\sum_{m=0,1,2\dots} \mathcal{J}_{m,n} \phi_m^{(0)}(y)\phi_0^{(0)^*}(y) \r].
\label{<l_t(y)>_st-series}
\end{align}
In fig.~\ref{<l_t(y)>_st-validation}, we have numerically checked that this series indeed converges to $\mathcal{K}(y,\op{min}(x_0,y))$ (defined in Eq.~\eqref{P_s-hp}). The expression (given in Eq.~\eqref{eq:<l_t(y)>-harm-alt-calc}) for the mean local time unconditioned on survival is verified for two cases, $y<x_0$ and $y>x_0$, in fig.~\ref{<l_t(y)>-harm-alt-calc}.

\section{Conclusions}
In this paper, we have studied the statistical properties of the local time $\ell_t$ spent  by an OU particle at different locations in the presence and in the absence of an absorbing boundary. 
Using the Feynman-Kac formalism, the MGF $\mathcal{Q}(k,t|x_0)$ of $\ell_t$ is written as an integral of the propagator of a quantum particle, for which the Hamiltonian naturally appears as the sum of a bare Hamiltonian and a perturbation. 
Exploiting this structure, we employ quantum perturbation theory to find the MGF as a power series in $k$, from which one can directly compute cumulants of $\ell_t$. 
This method is quite general, and can be applied for any functional $T_t[x(\tau)]$ of the form given in Eq.~\eqref{T_t[x]} and an arbitrary confining potential $V(x)$ (in Eq.~\eqref{eom}).
We find that for choices of $V(x)$ and $U(x)$ such that the Hamiltonian $\hat{H}_k$ with the effective potential $\mathcal{U}(x)+kU(x)$ in Eq.~\eqref{H_p} has a non-degenerate and isolated ground state, all the cumulants of $T_t[x(\tau)]$ increase linearly with time in the leading order. This observation leads us to 
predict a Gaussian distribution for the path functional, when shifted by its mean and scaled by its standard deviation. 
Since the Gaussian form of the distribution describes only the typical fluctuations, we study large deviations for large $t$ using saddle point calculations. Such a calculation indicates the existence of a Large Deviation Function $\mathbb{H}(w)$ which would describe the large fluctuations.
For the path functional $\ell_t(y)$, we find different asymptotic behaviours of $\mathbb{H}(w)$.

In the second part of the paper, we study statistical properties of $\ell_t(y)$ in the presence of an absorbing boundary, but not conditioned on survival. In this case, we have found that the distribution of $\ell_t(y)$ approaches a stationary distribution at large $t$, as expected from the theory of first passage functionals \cite{Majumdar2005}. In addition, we have shown how it approaches the stationary state by studying the MGF and computing the mean $\ell_t(y)$. For this, we have developed a general formalism which shows how the 
distribution approaches the stationary state. In particular, we demonstrate this approach explicitly by computing the mean local time for two cases --- a free particle and an OU particle.

We believe our results would be of interest to a broad community of physicists  working in the areas of stochastic processes, biological processes and chemical kinetics.
Our study can be naturally extended in several directions. 
For example, recently, there has been a lot of interest in active run-and-tumble particles which model the motion of a bacterium at the basic level.
For such particles, the problem of computing the MGF can be mapped to the evaluation of the eigenvalues and eigenfunctions of a Hamiltonian similar to that of a relativistic quantum particle \cite{Balakrishnan2005,Dhar2019}. It would be interesting to see how our results get modified for run-and-tumble particles. Another direction would be to allow the position of the absorbing boundary to vary.
It has been shown that in such scenario, the spectrum of the effective quantum potential undergoes a gap closing transition which manifests as a freezing transition in the barrier crossing problem \cite{Sabhapandit2020}. It would be interesting  to see how our results modify under such a transition.

\section*{Acknowledgement}
AK would like to acknowledge very stimulating discussions with Alain Comtet from which this project has originated. AK would also like to thank Prashant Singh for careful reading of the draft and for pointing out important mistakes. We  thank Hugo Touchette for helpful suggestions and for drawing our attention to important references. 
We also thank Alja\v{z} Godec for pointing out relevant references.
AK acknowledges support of the Department of Atomic Energy, Government of India, under project no.12-R\&D-TFR-5.10-1100. AK acknowledges support from DST, Government of India grant under project No. ECR/2017/000634. GK acknowledges the Long-Term Visiting Students Program through which he visited ICTS, Bangalore, where most parts of the project were carried out.

\newpage
\appendix
\section{}
\label{higher-order-terms}
Here, we provide the expressions for terms like $\lambda_n^{(0)},~\lambda_n^{(1)}, \dots$; $g_n^{(0)},~g_n^{(1)},\dots$; and $f_n^{(0)}(x_0),~f_n^{(1)}(y|x_0),\dots$ appearing in Eq.~\eqref{Q_k_expand}.
\begin{align}
\lambda_n^{(1)}(y) ={}&  \int dx\, \psi_n^{(0)^*}(x)\delta(x-y)\psi_n^{(0)}(x) = |\psi_n^{(0)}(y)|^2  \label{lam_n^1}\\
\lambda_n^{(2)}(y) ={}&  \sum_{m\neq n} \frac{\left \vert \psi_m^{(0)^*}(y)\psi_n^{(0)}(y) \right \vert ^2}{\lambda_n^{(0)}-\lambda_m^{(0)}} \label{lam_n^2}\\
\lambda_n^{(1,1)}(y,z) ={}& \sum_{m\neq n} \frac{[\psi_m^{(0)^*}(y)\psi_n^{(0)}(y)\psi_m^{(0)}(z)\psi_n^{(0)^*}(z)  + 
\psi_m^{(0)}(y)\psi_n^{(0)^*}(y)\psi_m^{(0)^*}(z)\psi_n^{(0)}(z)]}{\lambda_n^{(0)}-\lambda_m^{(0)}} \label{lam_n^11}
\end{align}
\begin{align}
f_n^{(0)}(x_0) ={}& \e^{\frac{V(x_0)}{2D}} { \psi_{n}^{(0)^*}}(x_0) \label{f_n^0}\\
f_n^{(1)}(y|x_0) ={}& \sum_{m\neq n} \frac{{ \psi_m^{(0)}}(y){ \psi_n^{(0)^*}}(y)}{\lambda_n^{(0)}-\lambda_m^{(0)}} ~ \e^{\frac{V(x_0)}{2D}} { \psi_{m}^{(0)^*}}(x_0)  \label{f_n^1} \\
\begin{split}
f_n^{(2)}(y|x_0) ={}& \sum_{m\neq n} \left [\sum_{p\neq n} \frac{\left[{ \psi_m^{(0)}}(y){ \psi_n^{(0)^*}}(y)  \left \vert \psi_p^{(0)}(y)\right \vert^2  \right ]}{\left(\lambda_n^{(0)}-\lambda_m^{(0)}\right)\left(\lambda_n^{(0)}-\lambda_p^{(0)}\right)} \right.  \\ 
& \qquad \qquad \qquad \qquad
\left.- \frac{\left[{ \psi_m^{(0)}}(y){ \psi_n^{(0)^*}}(y)  \left \vert \psi_n^{(0)}(y)\right \vert^2 \right]}{\left(\lambda_n^{(0)}-\lambda_m^{(0)}\right)^2}\right ]  { \psi_{m}^{(0)^*}}(x_0) \e^{\frac{V(x_0)}{2D}}
\end{split}
 \label{f_n^2} \\
f_n^{(1,1)}(y,z|x_0) ={}&  \sum_{m\neq n} \Bigg[\sum_{p\neq n} \frac{\left [{ \psi_m^{(0)}}(y)  { \psi_p^{(0)*}}(y) { \psi_p^{(0)}}(z) { \psi_n^{(0)^*}}({ z}) + { \psi_p^{(0)}}(y)  { \psi_n^{(0)^*}}(y) { \psi_m^{(0)}}(z) { \psi_p^{(0)^*}}({ z})\right ]}{\left(\lambda_n^{(0)}-\lambda_m^{(0)}\right)\left(\lambda_n^{(0)}-\lambda_p^{(0)}\right)}  \nonumber \\ 
& - \frac{\left[{ \psi_m^{(0)}}(y) { \psi_n^{(0)^*}}(y)\left \vert \psi_n^{(0)}(z)\right \vert^2  +{ \psi_m^{(0)}}(z) { \psi_n^{(0)^*}}(z)\left \vert \psi_n^{(0)}(y)\right \vert^2 \right]}{\left(\lambda_n^{(0)}-\lambda_m^{(0)}\right)^2}  \Bigg]  { \psi_{m}^{(0)^*}}(x_0) \e^{\frac{V(x_0)}{2D}}
 \label{f_n^11} 
\end{align}
where $V(x)=\kappa x^2/2$. Similarly from Eq.~\eqref{mcalQ} we get
\begin{align} 
 g_n^{(0)} ={}& \int dx\, \e^{-\frac{V(x)}{2D}} \psi_{n}^{(0)}(x), \label{g_n^0}\\
g_n^{(1)}(y) ={}&  \int dx\, \e^{-\frac{V(x)}{2D}}    \sum_{m\neq n} \frac{\psi_m^{(0)^*}(y)\psi_n^{(0)}(y)}{\lambda_n^{(0)}-\lambda_m^{(0)}}\psi_{m}^{(0)}(x)  \label{g_n^1} \\
\begin{split}
g_n^{(2)}(y) 
={}& \int dx\, \e^{-\frac{V(x)}{2D}} \sum_{m\neq n}  \left[\sum_{p\neq n} \frac{\left [\psi_m^{(0)^*}(y)\psi_n^{(0)}(y)  \left \vert \psi_p^{(0)}(y)\right \vert^2  \right ]}{\left(\lambda_n^{(0)}-\lambda_m^{(0)}\right)\left(\lambda_n^{(0)}-\lambda_p^{(0)}\right)} \right. \\ 
&\qquad \qquad \qquad \qquad \qquad \qquad
\left.- \frac{\left[\psi_m^{(0)^*}(y)\psi_n^{(0)}(y)  \left \vert \psi_n^{(0)}(y)\right \vert^2 \right]}{\left(\lambda_n^{(0)}-\lambda_m^{(0)}\right)^2}\right] \psi_{m}^{(0)}(x)
\end{split}
\label{g_n^2} \\
\begin{split}
g_n^{(1,1)}(y,z)
={}& \int dx\, \e^{-\frac{V(x)}{2D}} \sum_{m\neq n} \custombracket{[}{3em}{.} \sum_{p\neq n} \frac{ \psi_m^{(0)^*}(y)  \psi_p^{(0)}(y) \psi_p^{(0)^*}(z) \psi_n^{(0)}({ z}) }{\left(\lambda_n^{(0)}-\lambda_m^{(0)}\right)\left(\lambda_n^{(0)}-\lambda_p^{(0)}\right)}   \\ 
& \phantom{mmmm} + \sum_{p\neq n} \frac{\psi_p^{(0)^*}(y)  \psi_n^{(0)}(y) \psi_m^{(0)^*}(z) \psi_p^{(0)}({ z}) }{\left(\lambda_n^{(0)}-\lambda_m^{(0)}\right)\left(\lambda_n^{(0)}-\lambda_p^{(0)}\right)}  \\
& \phantom{mmmm} -
 \frac{\left[\psi_m^{(0)^*}(y) \psi_n^{(0)}(y)\left \vert \psi_n^{(0)}(z)\right \vert^2  +\psi_m^{(0)^*}(z) \psi_n^{(0)}(z)\left \vert \psi_n^{(0)}(y)\right \vert^2 \right]}{\left(\lambda_n^{(0)}-\lambda_m^{(0)}\right)^2} \custombracket{.}{3em}{]}
\psi_{m}^{(0)}(x)
\end{split}
\label{g_n^11}
\end{align}

\section{Functions (independent of time) appearing in Eqs.~\eqref{<l_t(y)l_t(z)>-har-FS} and \eqref{<ll>_a}}
\begin{align}
	\begin{split}
\mathcal{C}_{\langle \ell \ell \rangle}(y,z|x_0) &= \frac{1}{2D} \left[
		\frac{1}{\kappa \pi} \e^{-\frac{\kappa y^2}{2D}-\frac{\kappa z^2}{2D}} \sum_{\substack{m\ne 0\\l\ne 0}} \frac{H_m\!\left(\sqrt{\frac{\kappa}{2D} }y\right)H_m\!\left(\sqrt{\frac{\kappa}{2D} }x_0\right)}{m!\,m2^m} \frac{H_l\!\left(\sqrt{\frac{\kappa}{2D} }y\right)H_l\!\left(\sqrt{\frac{\kappa}{2D} }z\right)}{l!\,l2^l} \right.
		\\& - \frac{1}{\kappa \pi} \e^{-\frac{\kappa y^2}{2D}-\frac{\kappa z^2}{2D}}\sum_{m\ne 0}\frac{H_m\!\left(\sqrt{\frac{\kappa}{2D} }y\right)H_m\!\left(\sqrt{\frac{\kappa}{2D} }x_0\right)}{m!\,m^2 2^m} 
		\\& + \frac{1}{\kappa \pi} \e^{-\frac{\kappa y^2}{2D}-\frac{\kappa z^2}{2D}} \sum_{\substack{m\ne 0\\l\ne 0}} \frac{H_m\!\left(\sqrt{\frac{\kappa}{2D} }z\right)H_m\!\left(\sqrt{\frac{\kappa}{2D} }x_0\right)}{m!\,m2^m}\frac{H_l\!\left(\sqrt{\frac{\kappa}{2D} }z\right)H_l\!\left(\sqrt{\frac{\kappa}{2D} }y\right)}{l!\,l2^l} 
		\\& - \frac{1}{\kappa \pi} \e^{-\frac{\kappa z^2}{2D}-\frac{\kappa y^2}{2D}}\sum_{m\ne 0}\frac{H_m\!\left(\sqrt{\frac{\kappa}{2D} }z\right)H_m\!\left(\sqrt{\frac{\kappa}{2D} }x_0\right)}{m!\,m^2 2^m} 
		\\& \left.- \frac{1}{\kappa \pi} \e^{-\frac{\kappa y^2}{2D}-\frac{\kappa z^2}{2D}}\sum_{\substack{m\ne 0\\l\ne 0}} \frac{H_m\!\left(\sqrt{\frac{\kappa}{2D} }y\right)H_m\!\left(\sqrt{\frac{\kappa}{2D} }x_0\right)}{m!\,m 2^m} \frac{H_l\!\left(\sqrt{\frac{\kappa}{2D} }z\right)H_l\!\left(\sqrt{\frac{\kappa}{2D} }x_0\right)}{l!\,l 2^l} \right]
\end{split}
\label{C_(ll)} \\
	C_{\langle \ell \ell \rangle}^{(a)}(y,z|x_0) &= 
		\frac{y z^2\sqrt{\kappa} }{\pi \sqrt{2} D^{5/2}}
		\e^{-\frac{\kappa y^2}{2D}-\frac{\kappa z^2}{2D}}
		\sum_{\substack{p\ne 1\\\text{$p$ odd}}} 
		\frac{(-1)^{\frac{3p-1}{2}} p!!\,H_p\!\left(\sqrt{\frac{\kappa}{2D}}y\right)}{\l(1-p\r)^2 2^{\frac{p+1}{2}} p!\,p}
		 \nonumber \displaybreak[3]
		\\& - \frac{2}{\pi D^2}
		\frac{y z^2}{x_0} \e^{-\frac{\kappa y^2}{2D}-\frac{\kappa z^2}{2D}} 
		\sum_{\substack{m\ne 1\\\text{$m$ odd}}}
		\frac{H_m\!\left(\sqrt{\frac{\kappa}{2D}}y\right)H_m\!\left(\sqrt{\frac{\kappa}{2D}}x_0\right)}{2^mm! \l(1-m\r)^2}
		 \nonumber \displaybreak[3]
		\\& + \frac{z y^2\sqrt{\kappa} }{\pi \sqrt{2} D^{5/2}}
		\e^{-\frac{\kappa y^2}{2D}-\frac{\kappa z^2}{2D}}
		\sum_{\substack{p\ne 1\\\text{$p$ odd}}}
		\frac{(-1)^{\frac{3p-1}{2}} p!!\,H_p\!\left(\sqrt{\frac{\kappa}{2D}}z\right)}{2^{\frac{p+1}{2}} p!\,p \l(1-p\r)^2}
		 \nonumber \displaybreak[3]
		\\& - \frac{2}{\pi D^2}
		\frac{z y^2}{x_0} \e^{-\frac{\kappa y^2}{2D}-\frac{\kappa z^2}{2D}} 
		\sum_{\substack{m\ne 1\\\text{$m$ odd}}}
		\frac{H_m\!\left(\sqrt{\frac{\kappa }{2D}}z\right)H_m\!\left(\sqrt{\frac{\kappa}{2D}}x_0\right)}{2^mm! \l(1-m\r)^2}
		 \nonumber \displaybreak[3]
		\\& - \frac{z}{\pi\sqrt{2\kappa} D^{3/2}}
		\e^{-\frac{\kappa y^2}{2D}-\frac{\kappa z^2}{2D}}
		\sum_{\substack{p\ne 1\\\text{$p$ odd}}}
		\frac{(-1)^{\frac{3p-1}{2}} p!!\,H_p\!\left(\sqrt{\frac{\kappa}{2D}}y\right)}{2^{\frac{p+1}{2}} p!\,p \l(1-p\r)}
		\sum_{\substack{l\ne 1\\\text{$l$ odd}}} 
		\frac{H_l(\sqrt{\frac{\kappa}{2D}}y) H_l\!\left(\sqrt{\frac{\kappa}{2D}}z\right)}{\l(1-l\r) 2^ll!} 
		 \nonumber \displaybreak[3]
		\\& + \frac{2}{D \kappa\pi} \frac{z}{x_0}
		\e^{-\frac{\kappa y^2}{2D}-\frac{\kappa z^2}{2D}}
		\sum_{\substack{m\ne 1\\\text{$m$ odd}}}
		\frac{H_m\!\left(\sqrt{\frac{\kappa}{2D}}y_1\right)H_m\!\left(\sqrt{\frac{\kappa}{2D}}x_0\right)}{\l(1-m\r) 2^mm!}
		\sum_{\substack{l\ne 1\\\text{$l$ odd}}}
		\frac{H_l\!\left(\sqrt{\frac{\kappa}{2D}}y\right) H_l\!\left(\sqrt{\frac{\kappa}{2D}}z\right)}{\l(1-l\r) 2^ll!} 
		\nonumber\displaybreak[3]
		\\& - \frac{y}{\pi\sqrt{2\kappa} D^{3/2}}
		\e^{-\frac{\kappa y^2}{2D}-\frac{\kappa z^2}{2D}}
		\sum_{\substack{l\ne 1\\\text{$l$ odd}}} 
		\frac{H_l\!\left(\sqrt{\frac{\kappa}{2D}}z\right) H_l\!\left(\sqrt{\frac{\kappa}{2D}}y\right) }{\l(1-l\r) 2^ll!} 
		\sum_{\substack{p\ne 1\\\text{$p$ odd}}}
		\frac{(-1)^{\frac{3p-1}{2}} p!!\,H_p\!\left(\sqrt{\frac{\kappa}{2D}}z\right)}{2^{\frac{p+1}{2}} p!\,p \l(1-p\r)} 
		 \nonumber \displaybreak[3]
		\\& + \frac{2}{D \kappa\pi} \frac{y}{x_0}
		\e^{-\frac{\kappa y^2}{2D}-\frac{\kappa z^2}{2D}}
		\sum_{\substack{m\ne 1\\\text{$m$ odd}}}
		\frac{H_m\!\left(\sqrt{\frac{\kappa}{2D}}z\right)H_m(\sqrt{\frac{\kappa}{2D}}x_0)}{2^mm! \l(1-m\r)}
		\sum_{\substack{l\ne 1\\\text{$l$ odd}}}
		\frac{H_l\!\left(\sqrt{\frac{\kappa}{2D}}z\right) H_l\!\left(\sqrt{\frac{\kappa}{2D}}y\right)}{\l(1-l\r) 2^ll!}  
		\nonumber \displaybreak[3]
		\\& - \frac{y z}{\pi D^2} 
		\e^{-\frac{\kappa y^2}{2D}-\frac{\kappa z^2}{2D}}
		\sum_{\substack{l\ne 1\\\text{$l$ odd}}} 
		\frac{(-1)^{\frac{3l-1}{2}} l!!\, H_l\!\left(\sqrt{\frac{\kappa}{2D}}y\right)}{(1-l) 2^{\frac{l+1}{2}} l!\,l} 
		\sum_{\substack{m\ne 1\\\text{$m$ odd}}} 
		\frac{(-1)^{\frac{3m-1}{2}} m!!\, H_m\!\left(\sqrt{\frac{\kappa}{2D}}z\right)}{(1-m) 2^{\frac{m+1}{2}} m!\,m}
		 \nonumber \displaybreak[3]
		\\& - \frac{2}{D \kappa\pi}
		\frac{y z}{x_0^2} \e^{-\frac{\kappa y^2}{2D}-\frac{\kappa z^2}{2D}}
		\sum_{\substack{m\ne 1\\\text{$m$ odd}}}
		\frac{H_m\!\left(\sqrt{\frac{\kappa}{2D}}x_0\right)H_m\!\left(\sqrt{\frac{\kappa}{2D}}y\right)}{2^mm!\,(1-m)} 
		\sum_{\substack{m\ne 1\\\text{$m$ odd}}}
		\frac{H_m\!\left(\sqrt{\frac{\kappa}{2D}}x_0\right)H_m\!\left(\sqrt{\frac{\kappa}{2D}}z\right)}{2^mm!\,(1-m)}
		\label{C_(ll)^a} 
\end{align}

\section{Derivation of Eq.~\eqref{diff-eq-tildeP_ab}}
\label{deri-tildeP_a}
In this section, we derive Eq.~\eqref{diff-eq-tildeP_ab}. Here, we present the derivation for a general functional $T_t[x(\tau)]$ of the form given in Eq.~\eqref{T_t[x]}. 
The Laplace transform of the joint distribution $P_a(T,t_f \given x_0)$ of $T_t[x(\tau)] = \int_0^{t_f} U(x)\d\tau$ and the first-passage time $t_f = \int_0^{t_f}\d\tau$ is given by 

\begin{align}
	\begin{split}
		\LaplaceTransform{P}_a(k,q \given x_0) ={}& \int_0^\infty\d t_f \int_0^\infty \d T\, \e^{-kT}e^{-q t_f}P_a(T,t_f \given x_0)
	\end{split}
	\\\begin{split}
		={}& \mean{ \e^{-kT} \e^{-q t_f} }
	\end{split}
	\\\begin{split}
		={}& \mean{ \e^{\int_0^{t_f} \d\tau\, \l(- k U(x) - q\r)} }
	\end{split}
\end{align}
We split the interval $(0,t_f)$ into $(0,\Delta\tau) + (\Delta\tau,t_f)$.
One can then write
\begin{align}
	\begin{split}
		\LaplaceTransform{P}_a(k,q \given x_0) ={}& \mean{ \e^{\int_0^{t_f} \d\tau\, \l(- k U(x) - q\r)} }
	\end{split}
	\\\begin{split}
		={}& \mean{ \e^{ \Delta\tau \l(- k U(x) - q\r) } \LaplaceTransform{P}_a(k,q\given x_0+\Delta x) }_{\Delta x}
	\end{split}
\end{align}
where we average over all possible values of $\Delta x$, the distance the particle travels in the interval $(0,\Delta\tau)$.
\begin{align}
	\begin{split}
		\LaplaceTransform{P}_a(k,q \given x_0) ={}& \mean{ \e^{ \Delta\tau \l(- k U(x_0) - q\r) } \LaplaceTransform{P}_a(k,q\given x_0+\Delta x) }_{\Delta x}
	\end{split}
	\\\begin{split}
		={}& \mean{ \Bigg(1 - \Delta\tau \l(k U(x_0) + q\r)\Bigg) \l(\LaplaceTransform{P}_a + \frac{\dh \LaplaceTransform{P}_a}{\dh x_0} \Delta x + \frac{1}{2}\frac{\dh^2 \LaplaceTransform{P}_a}{\dh x_0^2} \Delta x^2\r)  }_{\Delta x}
	\end{split}
	\\\begin{split}
		={}& \widetilde{P}_a - \Delta \tau (k U(x_0)+q)~\widetilde{P}_a +\frac{\dh \LaplaceTransform{P}_a}{\dh x_0} \mean{\Delta x} + 
		\frac{1}{2}\frac{\dh^2 \LaplaceTransform{P}_a}{\dh x_0^2} \mean{\Delta x^2}
	\end{split}
	\label{tilde_P_a-ex}
\end{align}
From Eq.~\eqref{eom}, we have $\mean{\Delta x}=-V'(x_0)\Delta \tau$ and $\mean{\Delta x^2} = 2 D\Delta \tau +O(\Delta \tau^2)$. Inserting these in Eq.~\eqref{tilde_P_a-ex} and taking the $\Delta \tau \to 0$ limit, we get 
\begin{equation}
	D\frac{\dh^2 }{\dh x_0^2}\LaplaceTransform{P}_a(k,q|x_0)
	- V'(x_0)\frac{\dh }{\dh x_0} \LaplaceTransform{P}_a(k,q|x_0)
	-(k U(x_0)+q) \LaplaceTransform{P}_a(k,q|x_0) = 0
	\label{diff-eq-tildeP_a}
\end{equation}
The boundary conditions are
\begin{align}
	\lim_{x_0\to0} \LaplaceTransform{P}_a(k,q\given x_0) &= 1 \label{tildeP_a-bc-1-a}
	\\ \lim_{x_0\to\infty} \LaplaceTransform{P}_a(k,q\given x_0) &= 0 \label{tildeP_a-bc-2-a}
\end{align}
For the local time $\ell_t(y)$, one needs to put $U(x_0)=\delta(x_0-y)$ in Eq.~\eqref{diff-eq-tildeP_a}. 

\section{Some useful identities and relations}
\label{relations}
\begin{align}
\begin{split}
\mathscr{U}\l(-\frac{1}{2},z\r) &= \e^{-\frac{z^2}{4}} ,~~ \mathscr{U}\l(-\frac{1}{2},0\r) = 1 \\
\mathscr{V}\l(-\frac{1}{2},0\r) &= 0 \,,~~ \mathscr{V}\l(\frac{1}{2},z\r) = \frac{1}{\sqrt{\pi}} \e^{\frac{z^2}{4}} ,~~\mathscr{V}\l(-\frac{1}{2},z\r) = \e^{-\frac{z^2}{4}}\text{erfi}\l(\frac{v_0}{\sqrt{2}}\r) 
 \\
\text{erfi}(x)& =\frac{2}{\sqrt{\pi}}\int_0^x dy\, \e^{y^2} \\
\frac{d \mathscr{U}(a,z)}{dz} &= -\frac{z}{2} \mathscr{U}(a,z) - \l(a+\frac{1}{2} \r)\mathscr{U}(a+1,z) \\
\frac{d \mathscr{V}(a,z)}{dz} &= \mathscr{V}(a+1,z)-\frac{z}{2} \mathscr{V}(a,z) \\
\mathscr{U}(a,z) &\frac{d \mathscr{V}(a,z)}{dz} - \mathscr{V}(a,z) \frac{d \mathscr{U}(a,z)}{dz}  = 
\mathscr{U}(a,z) \,\mathscr{V}(a+1,z) +\l(a+\frac{1}{2} \r) \mathscr{V}(a,z) \,\mathscr{U}(a+1,z)
\end{split}
\label{eq:relations}
\end{align}

\section{Computation of $Q(k,t|x_0)$ for free particle}
\label{free-particle}
One can, in principle, compute the MGF $Q(k,t|x_0)$ using the perturbation expansion method given in sec.~\ref{MGF}. However, since the eigenspectrum is continuous for $V(x)=0$, the perturbation method is not especially advantageous  
over the usual backward Fokker-Planck method described in \cite{Majumdar2005}. Below, we use the latter method 
to compute $Q(k,t|x_0)$. The backward Fokker-Planck equation satisfied by the Laplace transformed MGF $\widetilde{Q}(k,q|x_0)=\int_0^\infty dt\, \e^{-qt} Q(k,t|x_0)$ is 
\begin{equation}
	D \frac{\dh^2 \LaplaceTransform{Q}}{\dh x_0^2} = k \,\delta(x_0-y) \LaplaceTransform{Q} + q \LaplaceTransform{Q} - 1
	\label{tildeQ(k,s)-free}
\end{equation}
with the boundary conditions
\begin{align}
\begin{split}
	&\evalAt{\LaplaceTransform{Q}(k,q\given x_0)}_{x_0\to 0} = 0
	\\&\evalAt{\LaplaceTransform{Q}(k,q\given x_0)}_{x_0\to\infty} =1
	\end{split}
	\label{tildeQ(k,s)-bc-free}
\end{align}
Due to the presence of the delta function in Eq.~\eqref{tildeQ(k,s)-free}, we write its general solution as
\begin{equation}
	\LaplaceTransform{Q} = 
		\begin{dcases}
			C_1 \e^{x_0\sqrt{\frac{q}{D}}} + C_2 \e^{-x_0\sqrt{\frac{q}{D}}} + \frac{1}{q} &,\, x_0<y\\
			C_3 \e^{x_0\sqrt{\frac{q}{D}}} + C_4 \e^{-x_0\sqrt{\frac{q}{D}}} + \frac{1}{q} &,\, x_0\geq y
		\end{dcases}
\end{equation}
where $C_1,~C_2,~C_3$ and $C_4$ are constants. They can be determined from the boundary conditions in Eq.~\eqref{tildeQ(k,s)-bc-free}, along with the continuity of $\widetilde{Q}(k,q|x_0)$ and the discontinuity of its derivative 
(with respect to $x_0$) across $x_0=y$. These extra conditions appear due to the presence of the delta function in Eq.~\eqref{tildeQ(k,s)-free}. Finding the constants, we finally get 
\begin{equation}
\resizebox{\textwidth}{!}{$
\LaplaceTransform{Q}(k,q|x_0) =
\begin{dcases}
\frac{1}{q} +\frac{k}{q\sqrt{4 q D}}\l[\frac{\l(1+\frac{k}{\sqrt{qD}}\op{sinh}\!\l(\sqrt{\frac{q}{D}}y\r)\r)}{\l(1+\frac{k}{\sqrt{qD}} \e^{-\sqrt{\frac{q}{D}}y} \op{sinh}\!\l(\sqrt{\frac{q}{D}}y\r)\r)} \e^{-2y\sqrt{\frac{q}{D}}} - \e^{-y\sqrt{\frac{q}{D}}}\r] \e^{\sqrt{\frac{q}{D}}x_0}  \\
\phantom{\frac{1}{q}+{}} +\frac{1}{q}\l[\frac{k}{\sqrt{4 q D}} \e^{\sqrt{\frac{q}{D}}y} - \frac{\l(1+\frac{k}{\sqrt{4 q D}}\r)\l(1+\frac{k}{\sqrt{qD}} \op{sinh}\!\l(\sqrt{\frac{q}{D}}y\r)\r)}{\l(1+\frac{k}{\sqrt{qD}} \e^{-\sqrt{\frac{q}{D}}y} \op{sinh}\!\l(\sqrt{\frac{q}{D}}y\r)\r)} \r] \e^{-\sqrt{\frac{q}{D}}x_0} 
&;\, x_0<y \\
\\
\frac{1}{q} - \frac{\l(1+\frac{k}{\sqrt{qD}} \op{sinh}\!\l(\sqrt{\frac{q}{D}}y\r)\r)}{q\l(1+\frac{k}{\sqrt{qD}} \e^{-\sqrt{\frac{q}{D}}y} \op{sinh}\l(\sqrt{\frac{q}{D}}y\r)\r)} \e^{-\sqrt{\frac{q}{D}}x_0} 
&;\, x_0\geq y
\end{dcases} 
\label{eq:Qtilde_s_uc_free}
$}
\end{equation}
Performing the inverse Laplace transform with respect to $q$, one can get $Q(k,t|x_0)$.

\section{Computation of large-time survival unconditioned moments for a free particle}
\label{free-particle-hom}
By using
\begin{equation}
	\bra{z}\hat{h}_k\ket{z'} = k \,\delta(y-z) \,\delta(z-z') \,,
\end{equation}
we simplify Eq.~\eqref{mbb_Q-free-particle} as
\begin{align}
	\mathbb{Q}(k,t|x_0) ={}& 1 - k \int_0^t d\tau \bra{y} \e^{-\hat{H}_k \tau}\ket{x_0},
	\\={}& 1 -  k\int_0^\infty d\tau \bra{y} \e^{-\hat{H}_k \tau}\ket{x_0} + k\int_t^\infty d\tau \bra{y} \e^{-\hat{H}_k \tau}\ket{x_0},
	\\={}& \mathcal{C}_k(y,x_0) + k\,\mathcal{A}_k(y,x_0|t),
	\label{eq: appendix F mathbbQ 1}
\end{align}
where we have defined
\begin{align}
	\mathcal{A}_k(y,x_0|t) &\defn \int_t^\infty d\tau \bra{y} \e^{-\hat{H}_k \tau}\ket{x_0},
	\\ \mathcal{C}_k(y,x_0) &\defn 1 -  k\int_0^\infty d\tau \bra{y} \e^{-\hat{H}_k \tau}\ket{x_0} = 1 - k \,\mathcal{A}_k(y,x_0|0),
	\\ \mathcal{G}_k(y,x_0|\tau) &\defn \bra{y} \e^{-\hat{H}_k \tau}\ket{x_0} \,.
\end{align}
We denote the Laplace transform of $\mathcal{G}_k(y,x_0|\tau)$ as
\begin{equation}
	\widetilde{\mathcal{G}}_k(y,x_0|s) = \int_0^\infty \d\tau \e^{-s\tau} \mathcal{G}_k(y,x_0|\tau) \,.
\end{equation}
Since
\begin{equation}
	\frac{d\mathcal{A}_k(y,x_0|t)}{d t} = - \mathcal{G}_k(y,x_0|\tau) \,,
\end{equation}
we can write
\begin{equation}
	\widetilde{\mathcal{A}}_k(y,x_0|s) = \frac{ \mathcal{A}_k(y,x_0|0) - \widetilde{\mathcal{G}}_k(y,x_0|s) }{s} \,.
\end{equation}
Using the above, we take the Laplace transform of both sides of Eq.~\eqref{eq: appendix F mathbbQ 1} and write
\begin{equation}
	\widetilde{\mathbb{Q}}(k,s|x_0) = \frac{ 1 - k\,\widetilde{\mathcal{G}}_k(y,x_0|s) }{s} \label{eq: appendix F mathbbQ 2}.
\end{equation}
We now calculate $\widetilde{\mathcal{G}}_k(y,x_0|\tau)$.
\begin{equation}
	\widetilde{\mathcal{G}}_k(y,x_0|s) 
	= \int_0^\infty \e^{-s\tau} \bra{y} \e^{-\hat{H}_k \tau}\ket{x_0} \,d\tau
	= \bra{y} \frac{1}{\hat{H}_k + s} \ket{x_0}
	= \bra{y} \frac{1}{\hat{H}_u + s +\hat{h}_k} \ket{x_0},
\end{equation}
where $\hat{H}_u$ is the unperturbed Hamiltonian.
We now define an operator corresponding to $\hat{\widetilde{\mathcal{G}}_k}(s) = [\hat{H}_u + s +\hat{h}_k]^{-1}$, such that
\begin{equation}
	\widetilde{\mathcal{G}}_k(y,x_0|s) = \bra{y} \hat{\widetilde{\mathcal{G}}_k}(s) \ket{x_0}
\end{equation}
This operator can be simplified as
\begin{align}
	\hat{\widetilde{\mathcal{G}}_k}(s) &= \frac{1}{\hat{H}_0 + s +\hat{h}_k}
	\\&= \left( \hat{H}_0 + s \right)^{-1} \left[ 1 + \left( \hat{H}_0 + s \right)^{-1} \hat{h}_k \right]^{-1}
	\\&= \hat{\widetilde{\mathcal{G}}_0}(s) \left[ 1 - \left( \hat{H}_0 + s \right)^{-1}\hat{h}_k \left[ 1 + \left( \hat{H}_0 + s \right)^{-1} \hat{h}_k \right]^{-1} \right]
	\\&= \hat{\widetilde{\mathcal{G}}_0}(s) - \hat{\widetilde{\mathcal{G}}}_0(s) \,\hat{h}_k \,\hat{\widetilde{\mathcal{G}}_k}(s)
\end{align}
Using the above, we can write
\begin{align}
	\widetilde{\mathcal{G}}_k(y,x_0|s) &= \bra{y}\hat{\widetilde{\mathcal{G}}}_0(s)\ket{x_0} - \iint d z d z' \bra{y} \hat{\widetilde{\mathcal{G}}}_0(s) \ket{z}\bra{z} \hat{h}_k \ket{z'}\bra{z'} \hat{\widetilde{\mathcal{G}}}_k(s) \ket{x_0}
	\\&= \widetilde{\mathcal{G}}_0(y,x_0|s) - k \,\widetilde{\mathcal{G}}_0(y,y|s) \,\widetilde{\mathcal{G}}_k(y,x_0|s)
	\\\implies \widetilde{\mathcal{G}}_k(y,x_0|s) &= \frac{ \widetilde{\mathcal{G}}_0(y,x_0|s)}{1 + k\,\widetilde{\mathcal{G}}_0(y,y|s) }
\end{align}
Putting the above in Eq.~\eqref{eq: appendix F mathbbQ 2}, we obtain
\begin{align}
	\widetilde{\mathbb{Q}}(k,s|y,x_0) ={}& \frac{1}{s}\left( 1 - \frac{ k\,\widetilde{\mathcal{G}}_0(y,x_0|s)}{1 + k\,\widetilde{\mathcal{G}}_0(y,y|s) } \right)
	\\={}& \frac{1}{s}\left[ 1 - k\,\widetilde{\mathcal{G}}_0(y,x_0|s) \sum_{n=0}^\infty (-k^n) \left(\widetilde{\mathcal{G}}_0(y,y|s) \right)^n \right]
	\\={}& \frac{1}{s}\left[ 1 + \sum_{n=1}^\infty (-k^{n}) \,\widetilde{\mathcal{G}}_0(y,x_0|s) \left(\widetilde{\mathcal{G}}_0(y,y|s) \right)^{n-1} \right]
\end{align}
From the above, it is clear that the Laplace transform of the $n$-th moment of local time is given by
\begin{equation}
	\widetilde{ \left< \ell^n_t(y) \right> }(s) = \frac{n!}{s} \,\widetilde{\mathcal{G}}_0(y,x_0|s)\left(\widetilde{\mathcal{G}}_0(y,y|s) \right)^{n-1}
	\label{eq: laplace transformed n-th moment exact expression free particle survival unconditioned}
\end{equation}

To evaluate $\widetilde{\mathcal{G}}_0(y,x_0|s)$ (which we will denote as $\mathcal{G}_s(y)$), we note that this Green's function satisfies
\begin{equation}
	\frac{\dh\mathcal{G}_0(y,x_0|t)}{\dh t} = D\frac{\dh^2\mathcal{G}_0(y,x_0|t)}{\dh y^2}
	\quad\quad\quad\text{ with }\quad
	\begin{aligned}
		&\mathcal{G}_0(y,x_0|t\to 0) = \delta(y-x_0) \\
		&\mathcal{G}_0(y\to 0,x_0|t) = 0 \\
		&\mathcal{G}_0(y\to\infty,x_0|t) = 0
	\end{aligned}
\end{equation}
Taking the Laplace transform of both sides of the above ($t\to s$), we write
\begin{equation}
	D\frac{\dh^2}{\dh y^2} \widetilde{\mathcal{G}}_0(y,x_0|s)(y) = s\, \widetilde{\mathcal{G}}_0(y,x_0|s)(y) - \delta(y-x_0)
\end{equation}
The presence of the Dirac $\delta$ in the above means that one should solve the above separately for $y<x_0$ and $y>x_0$, and then patch the two regions together by requiring continuity at $y=x_0$, along with a condition on the difference between the slopes on both sides.
On performing these calculations, one obtains
\begin{equation}
	\widetilde{\mathcal{G}}_0(y,x_0|s) = 
	\begin{dcases}
		\frac{1}{D}\sqrt{\frac{D}{s}} \exp\!\left( -\sqrt{\frac{s}{D}} \,x_0\right)  \op{sinh}\!\left(\sqrt{\frac{s}{D}} \,y\right) \,, & 0\le y\le x_0 \\
		\frac{1}{D}\sqrt{\frac{D}{s}} \exp\!\left(-\sqrt{\frac{s}{D}} \,y\right) \op{sinh}\!\left(\sqrt{\frac{s}{D}} \,x_0\right) \,, & x_0 \le y
	\end{dcases}
\end{equation}

The large-time limit corresponds to the limit $s\to 0$, in which case we can expand the above in a series and write
\begin{equation}
	\widetilde{\mathcal{G}}_0(y,x_0|s) = \frac{\op{min}(y,x_0)}{D} - \frac{y x_0}{D\sqrt{D}} \sqrt{s} + O(s^{3/2})
	\label{eq: mathcalG series small s}
\end{equation}
From Eq.~\eqref{eq: laplace transformed n-th moment exact expression free particle survival unconditioned}, we see that taking the inverse Laplace transform of $\widetilde{\mathcal{G}}/s$ should give us the mean local time at large time.
The expression obtained is the same as given in Eq.~\eqref{mean-lt-fp}.

Similarly, the Laplace-transform of the second moment is given by
\begin{align}
	\widetilde{\left<l_t^2(y)\right>} &= \frac{2}{s} \,\widetilde{\mathcal{G}}_0(y,x_0|s) \,\widetilde{\mathcal{G}}_0(y,x_0|s),
	%
	%
	%
	\\ &\approx \frac{2y\op{min}(y,x_0)}{ s D^2} - \frac{2 y^2 \left( x_0 + \op{min}(y,x_0) \right)}{D^2\sqrt{s D}} + O(1). 
\end{align}
where we have used Eq.~\eqref{eq: mathcalG series small s}.
Taking the inverse Laplace transform of the above, we obtain the following expression for the second moment:
\begin{align}
\begin{split}
	\left<l_t^2(y)\right>_{t\gg 1} &\approx \frac{2y\op{min}(y,x_0)}{ D^2} - \frac{2 y^2 \left( x_0 + \op{min}(y,x_0) \right)}{D^2\sqrt{\pi D t}} + O(t^{-1}).
\end{split}
\label{loc-var-exp-free-BM}
\end{align}

\end{document}